\documentclass[12pt]{article}
\pdfoutput=1
\usepackage[a4paper,bottom=4.2cm,top=1cm,head=3cm,width=18cm,dvipdfm]{geometry}
\evensidemargin=\oddsidemargin

\usepackage[dotinlabels]{titletoc}
\usepackage{titlesec}
\usepackage{ulem}

\usepackage{xcolor}
\usepackage{mathrsfs}
\usepackage{amsmath}
\usepackage{amssymb}
\usepackage{bm}
\usepackage{amsfonts}
\usepackage{subfigure}
\usepackage{feynmp}
\usepackage{extarrows}
\usepackage{slashed}
\usepackage{graphicx}
\usepackage{dcolumn}
\usepackage{verbatim}
\usepackage{here}
\usepackage{multirow} 
\allowdisplaybreaks[4]
\usepackage{indentfirst}
\usepackage{isodateo}
\usepackage[low-sup]{subdepth}

\definecolor{gesfpurple}{rgb}{0.47,0.19,0.42}

\definecolor{gesflanse}{rgb}{0.00,0.50,0.50}

\definecolor{gesfblue}{rgb}{0.08,0.42,0.76}

\definecolor{gesfred}{rgb}{1,0,0}

\definecolor{gesfwhite}{rgb}{1,1,1}

\definecolor{gesfblack}{rgb}{0,0,0}

\numberwithin{equation}{section}
\renewcommand{\thefootnote}{\arabic{footnote}}
\usepackage{isodateo}

\usepackage[linktocpage=true]{hyperref}
\hypersetup{
	colorlinks=true,
	citecolor=blue,
	linkcolor=blue,
	urlcolor=blue,
	pdfauthor={},
	pdftitle={},
	pdfsubject={}
}

\graphicspath{{figs/}}

\newcommand{\be}{\begin{equation}}
\newcommand{\ee}{\end{equation}}
\newcommand{\bea}{\begin{eqnarray}}
\newcommand{\eea}{\end{eqnarray}}

\def\ede{\end{equation}}
\def\bga{\begin{aligned}}
\def\eda{\end{aligned}}
\newcommand{\beq}{\begin{equation}}
\newcommand{\eeq}{\end{equation}}
\newcommand{\bq}{\begin{equation}}
\newcommand{\eq}{\end{equation}}
\newcommand{\ba}{\begin{array}}
\newcommand{\ea}{\end{array}}
\newcommand{\beqa}{\begin{eqnarray}}
\newcommand{\eeqa}{\end{eqnarray}}
\newcommand{\beqs}{\begin{subequations}}
\newcommand{\eeqs}{\end{subequations}}

\newcommand{\fr}[2]{\mbox{$\frac{\,{#1}\,}{#2}$}}
\def\nn{\nonumber}

\def\({\left(}
\def\){\right)}

\def\leqq{\leqslant}

\def\End{\end{document}}

\def\d{\text{d}}
\def\ii{{\tt i}}
\def\over{\overline}

\def\be{\beta}

\def\O{\mathcal{O}}

\def\Dm{\Delta{m}}
\def\mX{m_{\!X}^{}}
\def\mXP{m_{\!X'}^{}}
\def\cut{\Lambda}
\def\vDM{v_{\text{DM}}^{}}
\def\vDMM{v_{\text{DM}0}^{}}
\def\ER{E_R^{}}

\def\End{\end{document}}

\begin{document}

 \thispagestyle{empty}
 \renewcommand{\thefootnote}{\fnsymbol{footnote}}
 \setcounter{footnote}{0}
 \titlelabel{\thetitle.\quad \hspace{-0.8em}}
\titlecontents{section}
              [1.5em]
              {\vspace{4mm} \large \bf}
              {\contentslabel{1em}}
              {\hspace*{-1em}}
              {\titlerule*[.5pc]{.}\contentspage}
\titlecontents{subsection}
              [3.5em]
              {\vspace{2mm}}
              {\contentslabel{1.8em}}
              {\hspace*{.3em}}
              {\titlerule*[.5pc]{.}\contentspage}
\titlecontents{subsubsection}
              [5.5em]
              {\vspace{2mm}}
              {\contentslabel{2.5em}}
              {\hspace*{.3em}}
              {\titlerule*[.5pc]{.}\contentspage}
\titlecontents{appendix}
              [1.5em]
              {\vspace{4mm} \large \bf}
              {\contentslabel{1em}}
              {\hspace*{-1em}}
              {\titlerule*[.5pc]{.}\contentspage}

\def\thisday{February 23, 2017}



\begin{center}
{\Large\bf
EFT Approach of Inelastic Dark Matter
\\[2mm]
for Xenon Electron Recoil Detection}

\vspace*{8mm}

{\sc\large Hong-Jian He},$^{a,b,c,}$\footnote{Email: hjhe@sjtu.edu.cn, hjhe@tsinghua.edu.cn}~
{\sc\large Yu-Chen Wang},$^{b}$\footnote{Email: wang-yc15@mails.tsinghua.edu.cn}~
{\sc\large Jiaming Zheng}\,$^{a}$\footnote{Email: zhengjm3@sjtu.edu.cn}

\vspace*{4mm}

$^a$\,Tsung-Dao Lee Institute $\&$ School of Physics and Astronomy, \\
Shanghai Key Laboratory for Particle Physics and Cosmology,\\
Shanghai Jiao Tong University, Shanghai 200240, China
\\[1.5mm]
$^b$\,Institute of Modern Physics and Department of Physics, \\
Tsinghua University, Beijing 100084, China
\\[1.5mm]
$^c$\,Center for High Energy Physics, Peking University, Beijing 100871, China

\end{center}

\vspace*{12mm}


\begin{abstract}
\baselineskip 17pt
\noindent
Measuring dark matter (DM) signals via electron recoil provides an important means
for direct detection of light DM particles. The recent XENON1T anomaly with electron
recoil energy around $\,E_R^{}\!=\!(2-3)$\,keV can be naturally explained by DM
inelastic scattering which injects energy to the recoiled electrons and gives a
narrow peak structure in the recoil spectrum.
We present an effective field theory (EFT) approach to
exothermic inelastic DM signals for the Xenon electron recoil detection.
For relatively heavy mediator, we fairly formulate the DM-lepton interactions
by effective contact operators with two DM fields $(X,\,X')$ and two leptons.
Using the XENON1T data, we fit the electron recoil spectrum and constrain the allowed
scalar DM mass-splitting as $\,2.1\,\text{keV}\!<\!\Delta m\!<\!3.3\,\text{keV}$
(95\%\,C.L.), with the best fit $\,\Dm \!=\! 2.8\,$keV. We analyze the relic
abundance produced by such effective DM-electron contact interaction. To provide both
the DM relic abundance and the XENON1T excess, we derive new constraints on the
DM mass and the UV cutoff scale of DM effective interactions. Finally, we study
possible UV completions for the effective DM-lepton contact interactions.
\\[5mm]
JCAP (2020), no.12, in Press [\,arXiv:2007.04963\,].
\end{abstract}

\newpage
\renewcommand{\thefootnote}{\arabic{footnote}}
\setcounter{footnote}{0}
\setcounter{page}{2}

\tableofcontents

\baselineskip 17.5pt

\vspace*{10mm}
\section{Introduction}
\vspace*{1.5mm}
\label{sec:intro}
\label{sec:1}


With the tremendous experimental efforts of searching dark matter (DM) particles ranging
from the underground up to the sky over the past thirty years,
the dark matter physics is expected to be approaching an exciting turning point.
Among various ground-based experiments, measuring the DM signal via electron recoil
provides an important means for directly detecting light DM particles.
The XENON collaboration\,\cite{Aprile:2020yad} has newly announced an excess of events
with low electron recoil energy around
$E_R^{}\!=\!(2\!-\!3)\,{\rm keV}$ \cite{Aprile:2020tmw}.
The XENON1T detector recorded 285 events over the range
$E_R^{}\!=\!(1\!-\!7)\,{\rm keV}$, in which the the expected background events
are $\,232\pm 15$ \cite{Aprile:2020tmw}.
This leads to a $3.5\sigma$ excess above the expected backgrounds.
Such an anomaly may be attributed to possible tritium $\beta$ decays
in the backgrounds\,\cite{Aprile:2020tmw,Robinson:2020gfu} or by new
physics beyond the Standard Model (SM). In the latter case, the XENON
collaboration also pointed out two simple possibilities\,\cite{Aprile:2020tmw}:
(i) solar neutrinos with a large magnetic dipole moment\,\cite{nuMDM}
and (ii) solar axions\,\cite{solar_axion} absorbed by the recoiled electrons.
However, both explanations have
severe tension with stellar cooling constraints\,\cite{DiLuzio:2020jjp}
such as those from the white dwarfs and globular clusters.
Absorption of other light DM particles\,\cite{ALP-DP},
such as axion-like-particles or dark photons can also give rise
to a narrow peak signal at low recoil energy.

\vspace*{1mm}

Another class of explanations for the excess have focused on the scattering
between DM and electrons in XENON1T. But,
for an elastic DM-electron scattering process, it was found\,\cite{Kannike:2020agf}
that the DM particle $X$ has to be
as fast as $0.05c$ with mass $m_{X}\!\gtrsim\! 0.1\,\text{MeV}$
in order to produce the desired electron recoil energy of $O(\text{keV})$,
where $\,c\,$ denotes the light velocity.
This is an order of magnitude faster than the local escape velocity
$\,v_{\text{esc}}^{}\!\!\sim\!\!10^{-3}c$\, from the Milky Way.
There are related attempts to realize such a boosted DM
component for explaining the XENON1T excess\,\cite{boosted}.
Various other attempts also newly appeared\,\cite{others}\cite{Inelastic}.

\vspace*{1mm}

In this work, we investigate an attractive resolution that the electron recoil
is induced by inelastic scattering\footnote{During the completion of
writing this paper, some related papers appeared which explored the
inelastic DM explanation of XENON1T excess via specific DM
models with very light vector mediators\,\cite{Inelastic}.
Our EFT formulation of the
DM sector differs from all these because we have relatively heavy mediator
(either scalar or vector)
which can be integrated out from our EFT of the DM-electron interactions
at low energies.}
of a heavier DM component $X'$ to a lighter component $X$\,.
The exothermic inelastic DM scattering was studied before for a different purpose
which considered the DM scattering with nuclei as an explanation to the DAMA/LIBRA
excess\,\cite{Graham:2010ca}.
For the present study, we fit the XENON1T data and find that
the inelastic DM-electron scattering
leads to a narrow peak in the recoil spectrum
for the DM mass-splitting $\,\Dm \!\simeq\! (2.1\!-\!3.3)$\,keV,
which is consistent with the XENON1T excess.
This also means that its crossing channel will generate
the DM annihilation $\,XX'\!\!\to\! e^-e^+$.
We present an effective field theory (EFT) approach to inelastic DM
signals for the Xenon electron recoil detection.
For relatively heavy mediator, we can fairly formulate the DM-electron interactions
by effective contact operators with two DM fields $(X,\,X')$ and two electrons.
We demonstrate that the DM relic abundance
can be determined by the freeze-out of this annihilation process.
We compute the lifetime of the heavier DM component and find that it can be much
longer than the age of the Universe due to the small mass-splitting
required to explain the XENON1T excess.

\vspace*{1mm}

The paper is organized as follows. In Section\,\ref{sec:2},
we present the gauge-invariant effective operators of dimension-6
which realize exothermic inelastic scattering between the DM and electrons.
We analyze the contributions of these operators to the electron recoil spectrum
and fit them with the XENON1T data.
With this we identify the allowed parameter space for the inelastic DM.
In Section\,\ref{sec:3}, we study the contributions of the inelastic DM
to the relic abundance and derive the constraints.
We further analyze the lifetime of the heavier DM component $X'$ and
other constraints by the collider experiments.
In Section\,\ref{sec:4},
we discuss three possible UV completions of these effective operators.
Finally, we conclude in Section\,\ref{sec:5}.
Appendix\,\ref{app:A} presents a proof of the independent operators for
the DM-lepton interactions which are used for the current EFT approach.

\vspace*{2mm}
\section{EFT Approach of Inelastic Dark Matter\\ and Xenon Electron Recoil}
\label{sec:2}
\vspace*{1mm}

In this section, we study the inelastic DM as a resolution
to the XENON1T anomaly.
We present an effective field theory (EFT) approach to inelastic DM
signals for the Xenon electron recoil detection.
For relatively heavy mediator, we formulate the DM-electron interactions
by effective contact operators with two DM fields $(X, X')$ and two electrons.
Then, we analyze the predicted electron recoil energy spectrum
and identify the allowed parameter space by fitting the XENON1T data
and imposing the bound from CMB measurements of the DM relic abundance.

\vspace*{1mm}
\subsection{Realizing Minimal Inelastic Dark Matter}
\label{sec:2.1}
\vspace*{1mm}

For the present study, we construct a minimal dark sector for inelastic DM,
including two light real scalar DM fields $X$ and $X'$,
with masses around $\mX ,\mXP =O(\text{GeV})\gg m_{e}^{}$ and
a small mass-difference
$\,\Dm\equiv m_{X'}^{}-m_{X}^{}\!=\!O(2-3)\,{\rm keV}$.\,
In our EFT construction,
the DM fields $(X,\,X')$ are the SM singlets and odd under a $\mathbb{Z}_2^{}$ parity,
and their interactions with SM fields are realized via gauge-invariant effective operators
of dimension-6. These include the contact quartic interactions between $(X,\,X')$ and
lepton pairs relevant to the current study. Depending on the type of the bilinear lepton fields
in the quartic interaction, we may assign both $(X,\,X')$ as $P$-even real scalars or one of them
as $P$-odd pseudo-scalar. In the latter case, $(X,\,X')$ can combine to form a
complex singlet scalar such as $\widehat{X}\!=\!(X'\!+\!\ii X)/\!\sqrt{2}$ (for $X$ being $P$-odd) or
$\widehat{X}\!=\!(X\!+\!\ii X')/\!\sqrt{2}$ (for $X'$ being $P$-odd).

\vspace*{1mm}

In the present study, we will consider the following inelastic scattering process of
the DM with electrons,
\begin{equation}
e^{-}+X'\,\longleftrightarrow\, e^{-}+X \,.
\label{eq:eX'->eX}
\end{equation}
For the typical local DM velocity, it is known\,\cite{Kannike:2020agf} that
the recoil energy of an elastic scattering between
the DM and electron is too low to explain the event excess in the electron
recoil energy spectrum around $(2\!-\!3)$\,keV
as newly observed by XENON1T\,\cite{Aprile:2020yad}.
Thus, we only consider the inelastic channel \eqref{eq:eX'->eX}
for analyzing the xenon electron recoil detection.
If a large proportion of the DM is made of $X'$, then its exothermic
inelastic scattering $\,e^{-}+X'\!\rightarrow e^{-}+X$\, releases an amount of energy
$\sim\!\Delta m$ to the kinetic energy of the final states. In the case of
$\mX,\mXP\!\gg m_e^{}$, most of these released energy
goes into the kinetic energy of the scattered electron and produces
a narrow peak in the electron recoil energy $\,E_R^{}\!\sim\!\Delta m$\,.
The reverse process $\,e^{-}\!+X\!\rightarrow e^{-}\!+X'$\,
is kinematically suppressed because the local DM particles in average are too slow
to overcome the energy barrier.

\vspace*{1mm}

Related to the scattering process \eqref{eq:eX'->eX}, we note that the heavier
DM particle $X'$ has the following decay channel induced by an electron-loop,
\begin{equation}
X' \rightarrow X+(\text{photons})\,.
\end{equation}
This is the dominant decay channel if $X$ and $X'$
do not couple directly to light neutrinos.
The number of photons emitted in the decay products depends
on details of the DM-electron interaction and the spin of the DM particles.
This process is extremely suppressed kinematically because of the
small mass-splitting $\,\Delta m\!\ll\! \mX$\,.
The lifetime of $X'$ can be much beyond
the age of the Universe, as we will show later.

\vspace*{1mm}

This inelastic DM sector can be realized consistently in the early Universe.
The DM particles were originally in chemical equilibrium
with electrons/positrons through the annihilation,
\begin{equation}
e^{+}+e^{-}\longleftrightarrow X+X'\,.
\end{equation}
As the Universe cools down, the above scattering became inefficient when
$\,T\!\lesssim\! \mX$\,
and the dark matter relic density
$\,n_{X}^{}\!+n_{X'}^{}\,$
is determined by the usual freeze-out mechanism.
But, the scattering process \eqref{eq:eX'->eX} and its counterpart with positron
are still operative because of the large $e^\pm$ abundance.
These keep $\,n_{X}^{}\!\!=\!n_{X'}^{}\,$ for
$\,T\!\gg\!\Delta m$\,.
The process $\,X'X'\!\leftrightarrow\!XX\,$
also maintains chemical equilibrium between $X$ and $X'$.
But, it is mainly controlled by the quartic coupling
$X^2X'^2$ and we assume it decouples earlier than DM scattering with $e^\pm$.
The electron kinetically decouples
from $X$ and $X'$ at a temperature
$\,T_{D}^{}\!\approx\! m_{e}^{}\!\gg\!\Delta m$\,.
Since then the density ratio of $X$ and $X'$ has been frozen as
$\,n_{X}^{}=n_{X'}^{}$.
So only half of the DM particles ($X'$) can contribute to the event excess
in the electron recoil energy spectrum observed by the XENON1T experiment.

\vspace*{1mm}

In the following, we quantitatively compute the contribution
of our inelastic DM to the electron recoil energy spectrum of XENON1T.
We present an EFT formulation for the DM-lepton interactions
by considering the parameter space with the mediator mass much larger than
the energy scale of the DM-electron scattering and the lepton mass,
i.e., $M_{\text{md}}^2\gg q_t^2,\,m_\ell^2$,
where $M_{\text{md}}^{}$ denotes the mediator mass, $m_\ell^{}$ the lepton mass,
and $q_t^{}$ the 4-momentum transfer between the DM and electron.
In this case the DM-lepton interaction reduces to an effective contact operator.
This is a reasonable EFT setup since the freeze-out of the DM density and
the DM-electron scattering in XENON1T detector both occur at energy scales
much below $O(\text{GeV})$ which we will identify as the mass scale of our inelastic DM.
We may also consider the case with the mediator mass above the electroweak symmetry
breaking scale. Thus, integrating out the mediator field, we can write down the following
effective Lagrangian for the DM-lepton interactions:
\beqa
\label{eq:Leff}
\mathcal{L}_{(6)}^{} = \sum_j\frac{c_j^{}}{\,\tilde{\cut}^2\,}\O_j^{}
= \sum_j\frac{\,\text{sign}(c_j^{})\,}{\cut_j^2}\O_j^{} \,,
\eeqa
where each dimensionless coefficient $c_j^{}$ is the product of mediator couplings with
the DM and with the leptons and has possible signs $\text{sign}(c_j^{})=\pm$\,.
In the current notation,  $c_j^{}$ can be defined as a real coupling before specifying
the form of the corresponding operator $\O_j^{}$.
The UV cutoff scale $\tilde{\cut}$ equals the mediator mass,
$\tilde{\cut}=M_{\text{med}}^{}$, and we can define the corresponding effective
UV cutoff scale  $\cut_j^{}=\tilde{\cut}/{|c_j^{}|^{1/2}}$
for each operator $\O_j^{}$.
In the effective Lagrangian \eqref{eq:O6}, we have the following
$SU(2)_L^{}\otimes U(1)_Y^{}$ gauge-invariant and $CP$-conserving dimension-6
effective operators for the DM-lepton interactions,
\beqs
\label{eq:O6}
\beqa
{\cal O}_{Sj}^{} &\!\!\!=\!\!\!&
(\bar{L}H\ell_R^{})
[X'X,\, X^2\!,\, {X'}^2] + \text{h.c.}
\,,
\label{eq:O_S}
\\[1.5mm]
{\cal O}_{V\!L}^{} &\!\!\!=\!\!\!&
(\bar{L}\gamma^{\mu}L)
(X'\partial_{\mu}X\!-\!X\partial_{\mu}^{}X')\,,\,
\label{eq:O_VL}
\\[1.5mm]
{\cal O}_{V\!R}^{} &\!\!\!=\!\!\!&
(\bar{\ell}_R^{}\gamma^{\mu}\ell_R^{})
(X'\partial_{\mu}X\!-\!X\partial_{\mu}^{}X')\,,\,
\label{eq:O_VR}
\eeqa
\eeqs
where $H$ denotes the SM Higgs doublet with its vacuum expectation value (VEV)
$\left<H\right>=(0, v)^T$,\,
$L=(\nu_L^{},\,\ell_L^{})^T$ is the left-handed lepton-doublet, and
$\,\ell =e,\mu,\tau\,$.\footnote{%
Here we consider that the left-handed neutrino $\nu_L^{}$ will obtain Majorana
mass, so the right-handed neutrino $\nu_R^{}$ is either very heavy and decouples
from the low energy EFT or is absent. The case of neutrinos being pure Dirac type
can be considered as well, but we find it does not affect our main conclusion.}
For the vector-type interactions \eqref{eq:O_VL}-\eqref{eq:O_VR}, we assign one
of the scalars $(X,\,X')$ as $P$-odd and the other one as $P$-even, so the
operators ${\cal O}_{V\!L}^{}$ and ${\cal O}_{V\!R}^{}$ conserve $CP$.
We consider the scalar-type interactions \eqref{eq:O_S} and the vector-type interactions
\eqref{eq:O_VL}-\eqref{eq:O_VR} as motivated by two different types of the
underlying UV theories, so we will take them as two independent effective model-setups
for our present study and analyze them separately.
In the broken phase, the scalar-type operators \eqref{eq:O_S}
provide the following dimension-5
operators relevant to DM-lepton interactions,
\beqa
{\cal O}_{Sj}^{(5)} &\!\!\!=\!\!\!&
v(\bar{\ell}\ell)X'X,\, v(\bar{\ell}\ell)X^2\!,\, v(\bar{\ell}\ell){X'}^2 .
\label{eq:O_SS}
\eeqa
For the vector-type operators \eqref{eq:O_VL}-\eqref{eq:O_VR}, we find that they
give the same contributions to the DM scattering and annihilation amplitudes.
We also note that for ${\cal O}_{V\!L}$ and ${\cal O}_{V\!R}$,
the asymmetric combination
$(X'\partial_{\mu}^{}X\!-\!X\partial_{\mu}^{}X')$
is unique because the vector-type operators with the other combination
$(X'\partial_{\mu}^{}X\!+\!X\partial_{\mu}^{}X')=\partial_{\mu}^{}(XX')$
can be converted to terms suppressed by the leptonic Yukawa couplings of the SM,
or, to terms with additional gauge fields which are irrelevant to the current study.
Besides, in Eqs.\eqref{eq:O_VL}-\eqref{eq:O_VR},
we could consider the operators with
their DM part replaced by the bilinear fields
$X\partial_{\mu}^{}X \,(\propto\!\partial_\mu^{}X^2)$ or
$X'\partial_{\mu}^{}X' \,(\propto\!\partial_\mu^{}{X'}^2)$.
But by the same reasoning, we can convert such operators to terms
suppressed by the SM leptonic Yukawa couplings,
or, to terms with additional gauge fields.
Finally, we note that there is another dimension-6 operator involving
$U(1)_Y^{}$ gauge field strength $B^{\mu\nu}$ and the DM fields,
$\,B^{\mu\nu}\partial_\mu X\partial_\nu^{}X'$.
Again it can be converted to operators
suppressed by the SM leptonic Yukawa couplings.
The details of the proof are presented in Appendix\,\ref{app:A}.
We will further discuss the possible UV completions of the above effective
operators in Section\,\ref{sec:4}.

\vspace*{1mm}
\subsection{EFT Approach of Inelastic DM for Xenon Electron Recoil}
\label{sec:2.2}
\vspace*{1mm}

In this subsection, we use the generic effective DM-electron interactions
\eqref{eq:O6}-\eqref{eq:O_SS} to analyze the electron recoil energy spectrum
and compare it with the new measurement of XENON1T\,\cite{Aprile:2020tmw}.
We will demonstrate that the effective interactions \eqref{eq:O6}-\eqref{eq:O_SS}
can realize the inelastic DM scattering
$\,X'\,e^{-}\!\!\to\! X\,e^{-}$\,
and neatly explain the observed XENON1T anomaly.

\vspace*{1mm}

For the current situation, we note that the DM masses ($\mX,\,\mXP$),
their mass-splitting ($\Delta m$), and the electron mass ($m_e^{}$)
should obey the relation
$\,\Delta m\!\ll\! m_e^{}\!\!\ll\! \mX$.
In natural unit, we have the local DM velocity $\,\vDM\!\sim\!10^{-3}$\,
and typical atomic electron velocity
$\,v_e^{}\!\sim\!\alpha\!\sim\! 10^{-2}$,
where $\alpha$ is the fine structure constant.
So we have the velocity relation
$\,v_{\text{DM}}^{}\!\ll\! v_{e}^{}\!\ll\! 1$\,.
Thus, for the inelastic scattering $\,X'\,e^{-}\!\!\to\! X\,e^{-}$,\,
we can express the electron recoil energy
to the leading order of $(\vDM,\, v_e^{})$ and $(\Delta m,\, m_e^{})$,
\beqa
\label{eq:ER}
\ER \,\simeq\, \Delta m\!\(\!1-\frac{\,\vDM\,}{v_{e}}\cos\theta_e^{}\!\)\! ,
\eeqa
where $\theta_e^{}$ is the scattering angle between the moving directions of
the final and initial state electrons.
In deriving the above formula, we have chosen the initial state electron
and $X'$ to move in parallel for simplicity of demonstration, but
the following analysis does not rely on this choice.
Because $\,v_{\text{DM}}^{}\!\ll\! v_{e}^{}$, Eq.\eqref{eq:ER} shows that
the recoil energy spectrum has to exhibit a narrow peak around
$\,\ER\approx\Delta m\,$,
which will be further spread by detector resolution.

\vspace*{1mm}

To analyze the electron recoil energy spectrum at XENON1T induced by the
inelastic DM scattering, we use the systematic treatment of
\cite{Essig:2011nj}-\cite{Roberts:2019chv}.
We parameterize the $X'e$ scattering cross section as
$\,\sigma_{\!Xe}^{}(q) \!=\! \over{\sigma}_{\!e}^{}\,|F_{X}^{}(q)|^{2}$,\,
where $\,q\!\equiv\!|\vec{q}\hspace*{0.5mm}|$\, is the size of transferred 3-momentum,
and $\,\over{\sigma}_{\!e}^{}\!\equiv\!\sigma_{\!Xe}^{}(q\!=\!0)$\,
is the scattering cross section evaluated at $\,q\!=\!0$\,.
The function $F_X^{}(q)$ is the DM form factor that captures the $q$
dependence of the cross section. We consider the DM mass range
$\,\mX \!\gg m_e^{}$\,. For the inelastic scattering $X'\,e^{-}\!\!\to\! X\,e^{-}$
induced by scalar-type contact interaction $\O_{S1}^{(5)}$ in Eq.\eqref{eq:O_SS},
we derive
\beqs
\label{eq:sigma-S}
\begin{align}
\label{eq:sigma-SS}
\over{\sigma}_{\!e}^{S} &
=\,\frac{m_e^{2}\,v^2}{\,4\pi m_X^2\Lambda_S^4\,} \,,
\\[1mm]
\label{eq:sigma-SFX}
|F_{X}^{S}(q)|^{2} & =\,\frac{\,m_{e}^{2}+q^2\!/4\,}{m_{e}^{2}}\,,
\end{align}
\eeqs
where $\cut_S^{}$ is the effective cutoff scale associated with the operator
$\O_{S1}^{}$ and $\O_{S1}^{(5)}$.\,
For vector-type contact interaction ${\cal O}_{V\!L}$ or ${\cal O}_{V\!R}$
in Eqs.\eqref{eq:O_VL}-\eqref{eq:O_VR}, we denote the associated cutoff scale
as $\,\cut_V^{}$ and deduce the following,
\beqs
\label{eq:sigma-V}
\begin{align}
\label{eq:sigma-VV}
\over{\sigma}_{\!e}^{V} & =\,
\frac{m_{e}^{2}}{\,4\pi\Lambda_{V}^4\,} \, ,
\\[1mm]
\label{eq:sigma-VFX}
|F_{X}^{V}(q)|^{2} & =\,
\frac{\,m_e^2-q^2\!/2\,}{m_{e}^{2}} \,.
\end{align}
\eeqs

Then, the velocity-averaged differential cross section is given by
\beqa
\frac{\,{\rm d}\langle\sigma_{\!Xe}^{}\vDM\rangle\,}{\,{\rm d}\ER\,}
=\,\frac{\over{\sigma}_{\!e}^{}}{\,2m_e^{}\,}\!\!
\int\!{\rm d}\vDM\frac{\,f(\vDM)\,}{\vDM}\!
\int_{q_{-}}^{q_{+}}\!\!\!{\rm d}q\,a_0^2\,q\,|F_X^{}(q)|^{2}K(\ER ,q)\, ,
\label{eq:DSigDE}
\eeqa
where $\,f(\vDM )\,$ is the local DM velocity distribution function,
normalized to $\int\!{\rm d}v_{\text{DM}}\,f(v_{\text{DM}})=1\,$.
We take $\,f(\vDM )\,$ as a pseudo-Maxwellian distribution,
\beqa
f(v) \,=\,
N_0^{}v^2 \exp\!\left[-(v\!-\!v_{\text{mean}}^{})^2/(2v_{\text{rms}}^2)\right],
\eeqa
where $N_0^{}$ is the normalization factor,
$\,v_{\text{mean}}^{}$ denotes the average velocity
$\,v_{\text{mean}}^{}\!\!=\!0.77\!\times\!10^{-3}$,\,
and $\,v_{\text{rms}}^{}$ is the local DM velocity dispersion
$\,v_{\text{rms}}^{}\!=\!0.73\!\times\!10^{-3}$ \cite{Freese:2012xd}.
In the above Eq.\eqref{eq:DSigDE}, $\,a_0^{}=1/(m_e^{}\alpha)$\,
is the Bohr radius, while
$K(E ,q)$\, is the atomic excitation factor.
We input $K(E ,q)$ from Fig.7 of \cite{Roberts:2019chv}
with $\,\ER\! =2\,\text{keV}$.
Most of our signal events have a recoil energy around
$\,\ER\!\sim\!\Delta m \!=\!(2-3)\,\text{keV}$.
For $\ER\!=\!(1-5)\,\text{keV}$,
the scattering happens dominantly with
electrons in the $3s$ shell\,\cite{Roberts:2016xfw}.
The function $K(E ,q)$ is independent of $E$ before it reaches the threshold of
the next quantum energy level.
So we can approximate
$\,K(E ,q)\!\simeq\! K(\Delta m,q)\!\simeq\! K(2{\rm keV}\!,q)$\,
for the calculation. The upper and lower limits of the $q$ integration
($\,q_-^{}\!\!\leqq\! q \!\leqq\! q_+^{}$\,)
is determined by the range which obeys the energy-momentum conservation,
\beqa
q^{2}\!-2q \mX \vDM \!\cos\hspace*{-0.4mm}\eta + 2\mX (\ER\!-\!\Delta m) \,=\,0 \,,
\eeqa
for any $\eta$\,, where
$\,\eta\,$ is the angle between the momentum-transfer $\,\vec{q}\,$
and the momentum $\,\vec{p}_i^{}\,$ of the incident DM particle $X'$.
Thus, we have
\begin{equation}
\label{eq:q+-}
\frac{q_{\pm}^{}}{\,\mX\,} \,=\,
\left|\vDM\pm
\sqrt{v_{\text{DM}}^2-2\left(\frac{E_{R}\!-\!\Delta m}{\mX}\right)\,}\,\right| .
\end{equation}
The electron recoil energy spectrum of the scattering events is given by
\beqa
\label{eq:dN/dE}
\frac{\d N}{\,\d \ER\,}\,\simeq\,
\frac{\,{\rm d}\langle\sigma_{\!Xe}^{}\vDM\rangle\,}{\,{\rm d}\ER\,}
\frac{\rho_{\text{DM}}^{}}{\,2m_{\!X'}^{}\,}N_T^{}\,\Delta t \,,
\eeqa
where the product
$\,N_T^{}\,\Delta t\simeq 4.2\times 10^{27}\!/
 \text{ton}\times 0.65\,\text{ton}\cdot\text{yr}$\,,
which gives the number of atoms $N_T^{}$ times the total exposure time
$\Delta t$ for the Science Run-1 (SR1)\,\cite{Aprile:2020tmw}.
The local DM density is
$\,\rho_{\text{DM}}^{}\!\simeq\! 0.3\,\text{GeV\!/cm}^{3}$.\,
We have used the condition
that the heavier component $X'$ makes up half of the dark matter relics.
We further incorporate the detector energy resolution
$\,\sigma_{\!d}^{}=0.5\,{\rm keV}$\,\cite{Aprile:2020yad}
into the recoil spectrum by convolving it with a normal distribution
$G(E,\sigma_{\!d}^{})$. The efficiency function $\eta(E)$ of the XENON1T detector
is given by Fig.\,2 of Ref.\,\cite{Aprile:2020tmw}.
Thus, we estimate the detected recoil energy spectrum as follows,
\beqa
\frac{{\rm d}N^{}_{\text{DT}}}{\,{\rm d}\ER\,} \,=\,
\eta(\ER )\!\!\int\!\!{\rm d}E'\,G(\ER\!-\!E',\sigma_{\!d}^{})
\frac{\,{\rm d}N(E')\,}{{\rm d}E'} \,,
\eeqa
where $N^{}_{\text{DT}}$ denotes the detected number of
the DM-electron scattering events.

\vspace*{1mm}

In the above formulation, we have three key quantities for describing the
DM-electron inelastic scattering: the DM mass $\mX$, the DM mass-splitting $\Dm$,
and the inelastic scattering cross section at $q^2=0$\,, which is
$\sigma_{\!Xe}^{}(q^2\!=\!0)=\over{\sigma}_{\!e}^{}$.\,
From Eqs.\eqref{eq:sigma-S}-\eqref{eq:sigma-V}, we see that in the expression of
the inelastic cross section
$\sigma_{\!Xe}^{}$, the part $\over{\sigma}_e^{}$ contains all the information
of the DM-electron interactions, especially the effective DM-electron coupling
$\cut_S^{-1}$ or $\cut_V^{-1}$ as defined in the dimension-6 effective operators
\eqref{eq:Leff}-\eqref{eq:O_SS}.
Also, the kinematic function $F_X^S(q)$ or $F_X^V(q)$ just extracts the
$q^2$-dependence of the inelastic cross section $\sigma_{\!Xe}^{}$, and
practically $\,F_X^{S,V}(q)\simeq 1\,$ holds well for the relevant region of the
$q$-integration \eqref{eq:DSigDE}. This is because the atomic excitation function $K(E,q)$ takes its peak value at $q\approx 0.04\,\text{MeV}\!\ll\!m_e^{}$ and falls off rapidly as $q$ deviates from this peak position.
Hence, we can make a {\it fairly model-independent fit} of the
inelastic DM parameters $(\mX,\,\Dm,\,\over{\sigma}_e^{})$
with the XENON1T data.

\vspace*{1mm}

\begin{figure}   
\centering
\includegraphics[width=8.6cm,height=7cm]{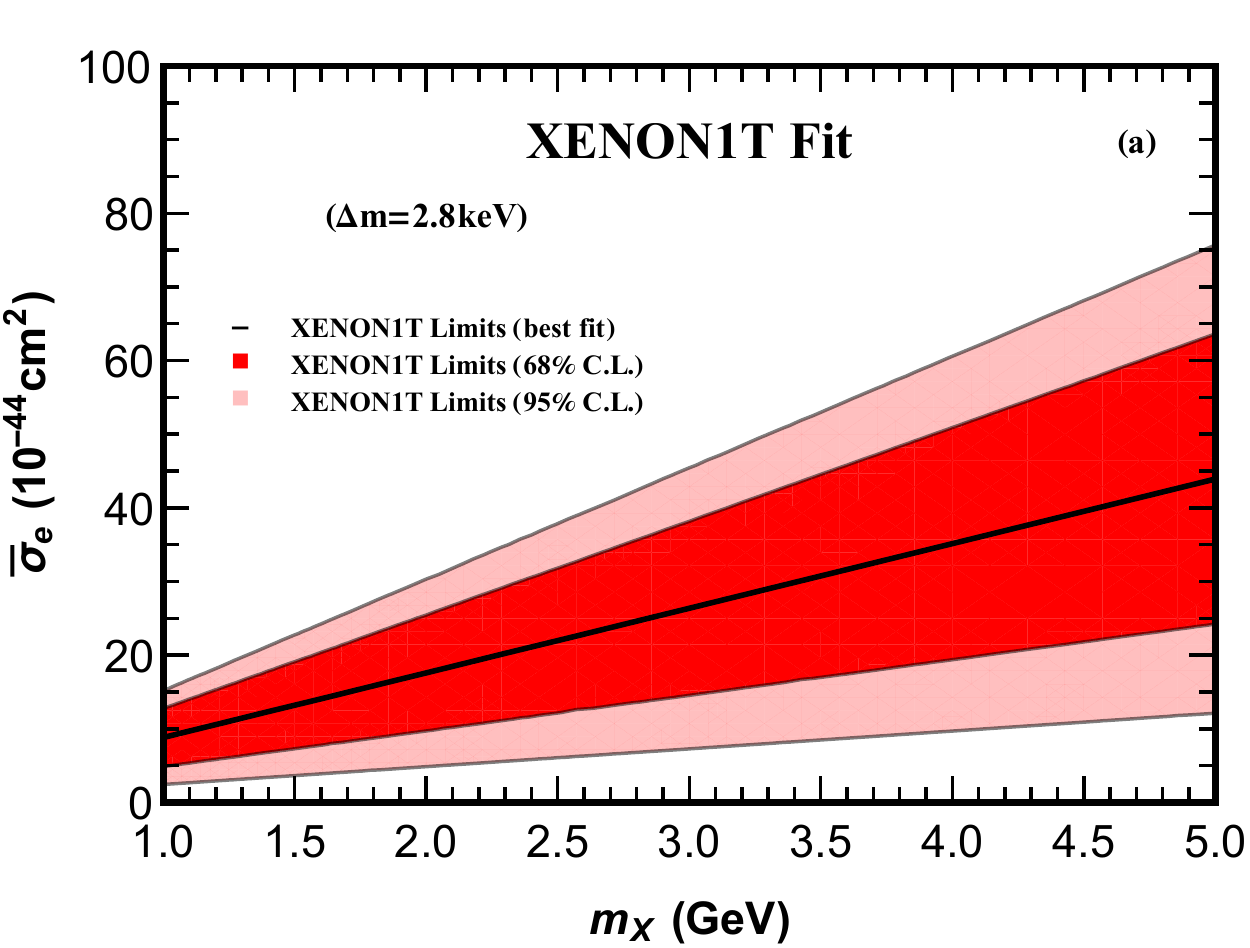}
\hspace*{-1.5mm}
\includegraphics[width=8.1cm,height=7cm]{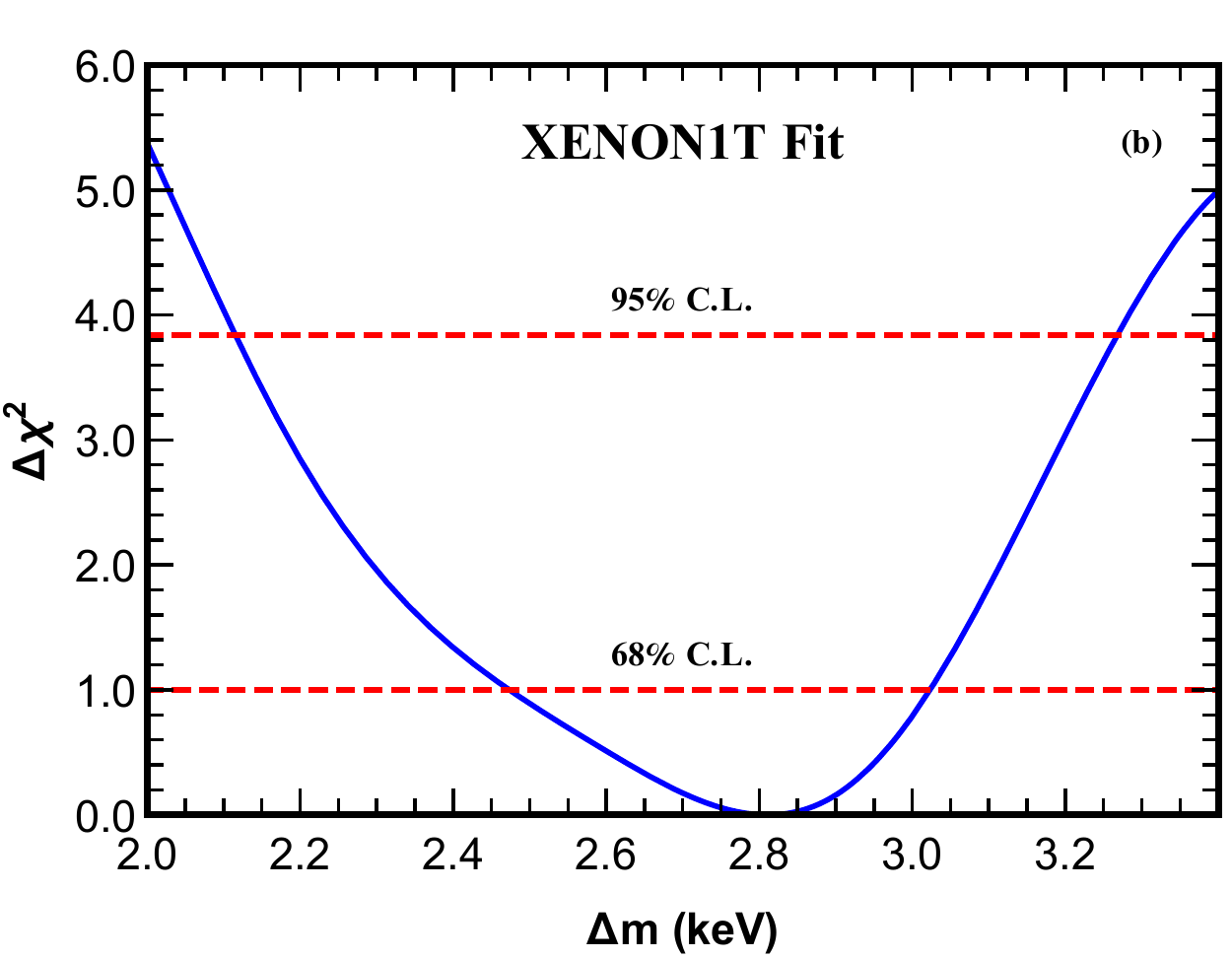}
\caption{\small
Fitting inelastic DM with XENON1T data.
The fit is performed by varying the parameters
$(\mX,\,\Dm,\,\over{\sigma}_e^{})$ simultaneously.
Plot-(a) presents the allowed parameter space in the
$(\mX ,\,\over{\sigma}_e^{})$ plane
for setting the DM mass-splitting to its best fit $\Dm=2.8$\,keV,
where the red and pink contours give the 68\% and 95\% confidence limits, respectively.
The black solid curve in the middle of the contours correspond to the best fit
of $(\mX,\,\over{\sigma}_e^{})$.
Plot-(b) shows the $\chi^2$ fit for $\Dm$, which gives the best fit of
$\Dm =2.8$\,keV, and the allowed ranges of
$\,\Dm =2.8^{+0.2}_{-0.3}$\,keV (68\%\,C.L.) and
$\,2.1\,\text{keV}\!<\! \Dm\!<\! 3.3\,\text{keV}$ (95\%\,C.L.).
\label{fig:1}
}
\end{figure}

With these, we present our fitting results in Fig.\,\ref{fig:1}.
Plot-(b) shows the $\chi^2$ fit for $\Dm$\,, which gives the best fit
of the DM mass-splitting
$\,\Dm \!=\!2.8$\,keV, and the allowed ranges:
$\,\Dm =2.8^{+0.2}_{-0.3}$\,keV (68\%\,C.L.) and
$\,2.1\,\text{keV}\!<\! \Dm\!<\! 3.3\,\text{keV}$ (95\%\,C.L.).
In Plot-(a), we present the allowed parameter space in the $\mX\!-\over{\sigma}_e^{}$ plane
by fixing the DM mass-splitting to its best fit $\,\Dm\!=\!2.8$\,keV,
where the red and pink contours correspond to the 68\%\,C.L.\ and 95\%\,C.L.\ limits, respectively.
The black solid curve in the middle of the contours corresponds to the best fit
of $(\mX,\,\over{\sigma}_e^{})$.
As we will show, given the general contour of
$(\mX,\,\over{\sigma}_e^{})$  in Fig.\,\ref{fig:1}(a),
we can further derive new bounds on the cutoff scale $\Lambda$ versus the DM mass
$\mX$ for each given type of contact DM-electron interactions.

\vspace*{1mm}

\begin{figure}
\centering
\includegraphics[width=11cm,height=8cm]{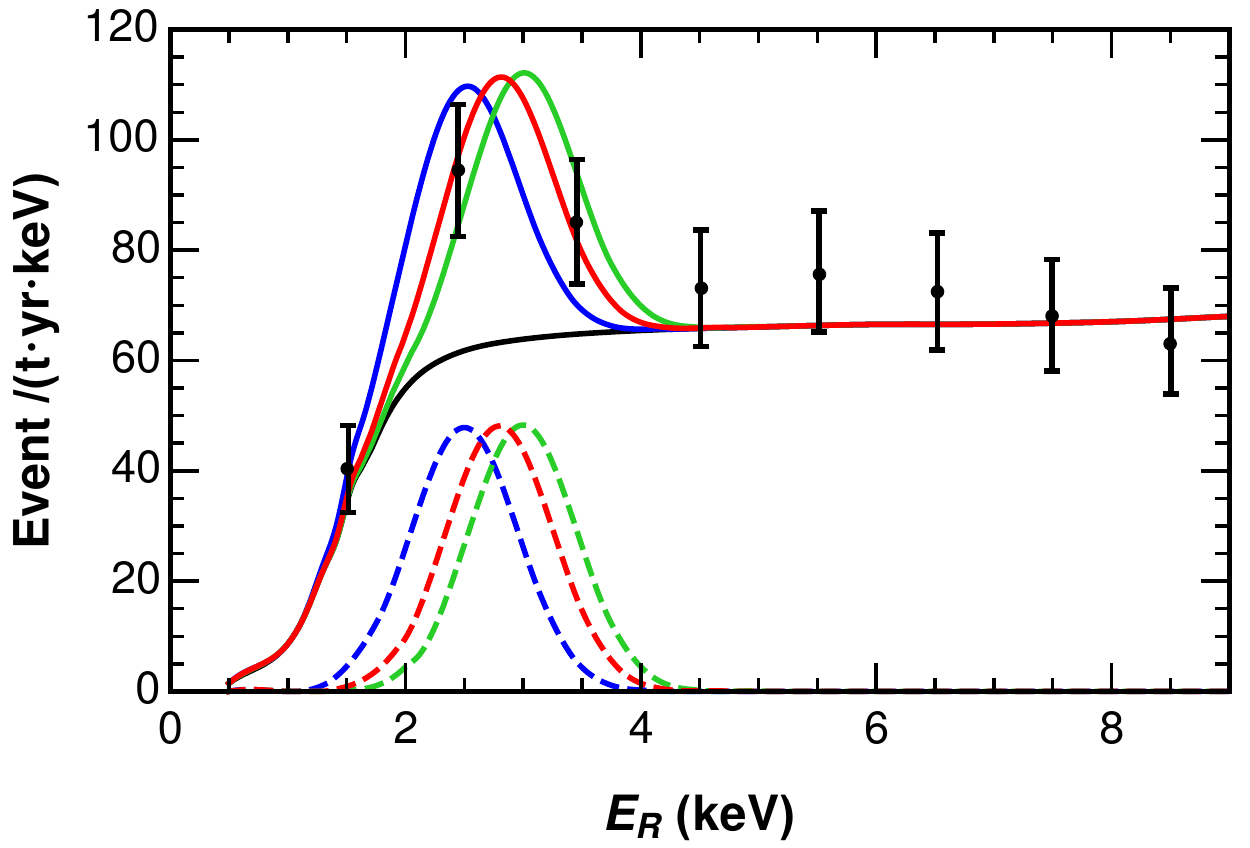}
\vspace*{-2mm}
\caption{\small
Prediction of inelastic DM for the electron recoil energy spectrum and the XENON1T data.
The data points with error bars correspond to the new measurement of XENON1T\,\cite{Aprile:2020tmw},
and the black solid curve shows the background contribution.
The (green, red, blue) solid curves include the inelastic DM contributions with
the DM mass-splitting $\,\Delta m\!=\!(2.5,\,2.8,\,3.0)$\,keV, respectively, whereas
the dashed (green, red, blue) curves correspond to the inelastic DM contributions alone.
We have input a sample cross-section/mass ratio
$\,\over{\sigma}_e^{}/\mX\! =8.8 \!\times\! 10^{-44}\text{cm}^2/$GeV,
which corresponds to the best fit of Fig.\,\ref{fig:1}(a).
\label{fig:signal}
\label{fig:2}
}
\end{figure}

Inspecting Eqs.\eqref{eq:DSigDE} and \eqref{eq:dN/dE}, we observe that
the information of the DM dynamics enters the recoil spectrum via the
ratio $\,\over{\sigma}_e^{}/\mXP$ (or equivalently, $\over{\sigma}_e^{}/\mX$)
for a given DM mass-splitting $\Dm$\,. The integral upper/lower limits $q_\pm^{}$
[cf.\ Eq.\eqref{eq:q+-}] also have dependence on $\mX$, but we find that
this effect is rather weak and practically negligible for the final result.
In Fig.\,\ref{fig:2}, we present the smeared electron recoil energy spectrum
for the sample input of cross-section/mass ratio
$\over{\sigma}_e^{}/\mX \!=\! 8.8 \!\times\! 10^{-44}\text{cm}^2/$GeV,
which corresponds to the best fit of Fig.\,\ref{fig:1}(a).
As will be shown below, this input satisfies the constraint of
the DM relic abundance.
The data points with error bars correspond to the new measurement by
the XENON1T collaboration and the black solid curve shows the background contribution
in the XENON1T detector, which are taken from Ref.\,\cite{Aprile:2020tmw}.

\vspace*{1mm}

With our generic EFT formulation of the inelastic DM, we have computed the
electron recoil energy spectrum for different DM mass-splittings
$\,\Delta m=(2.5,\,2.8,\,3.0)$keV, which are plotted as
(blue, red, green) dashed curves in Fig.\,\ref{fig:2}.
We sum these DM signal contributions with
the backgrounds (black solid curve) respectively, and plot them as the
(green, red, blue) solid curves in the same figure.
It shows that the case of $\Delta m=2.8\,$keV (red solid curve)
gives the best fit to the recoil spectrum measured by XENON1T.
Also, comparing the (blue, red, green) solid curves with different $\Delta m$ values,
we see that varying the $\Dm$ value has little effect on the height and width
of the recoil peak, but it does shifts the peak position in $\ER$\,.
We see that even after including the detector energy resolution the recoil peak still
 remains quite narrow, so the peak position in $\ER$ is fairly constrained by the
XENON1T data.

\vspace*{1mm}

Next, we can apply the general fit of Fig.\,\ref{fig:1}(a) to the case of
the scalar-type DM-electron interaction \eqref{eq:O_SS}
and to the case of the vector-type DM-electron \eqref{eq:O_VL}-\eqref{eq:O_VR},
respectively. With the fit of Fig.\,\ref{fig:1}(a), we can derive the allowed parameter
space in the $\mX \!-\cut_S^{}$ plane for the scalar-type DM-electron interaction,
and in the $\mX \!-\!\cut_V^{}$ plane for the vector-type DM-electron interaction.
This is practically equivalent to making a direct fit of XENON1T data
(under $\,\Dm =2.8\,$keV) in the $\mX \!-\!\cut_S^{}$ plane and in
the $\mX \!-\!\cut_V^{}$ plane, respectively.

\begin{figure}
\centering
\includegraphics[width=8.3cm,height=7cm]{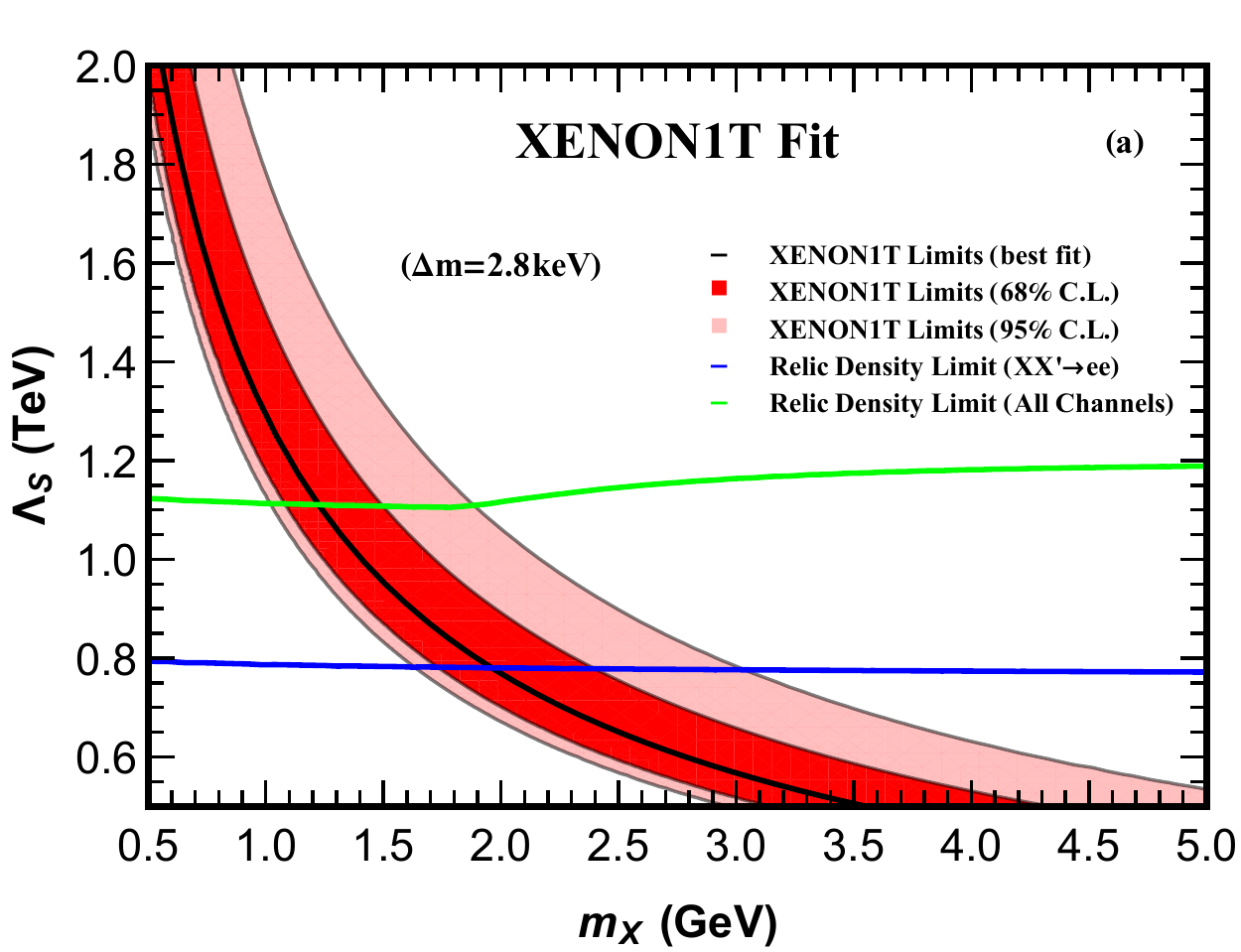}
\hspace*{-3mm}
\includegraphics[width=8.3cm,height=7cm]{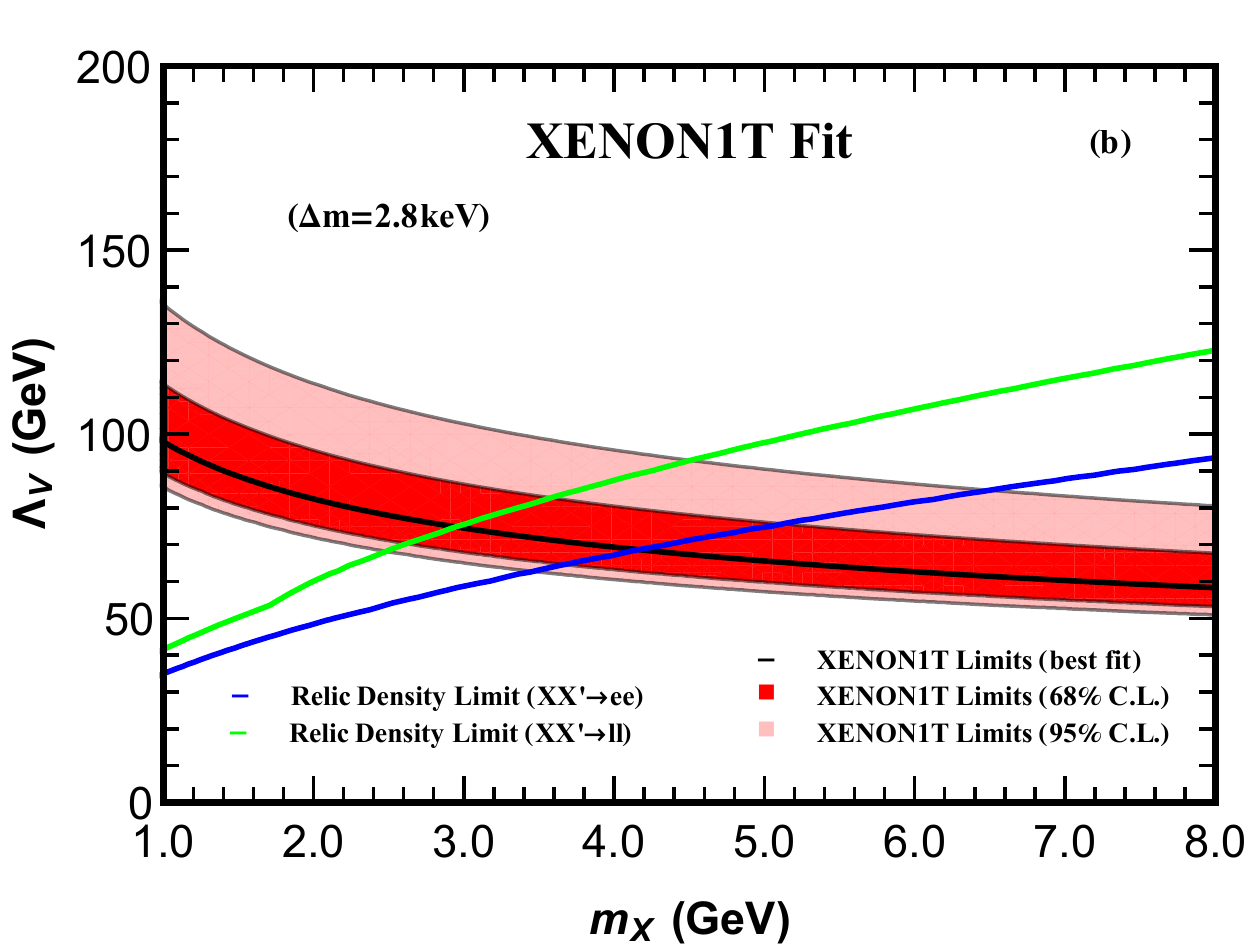}
\caption{\small
Bounds on the parameter space $(\mX,\,\Lambda)$ of the inelastic DM,
as derived from XENON1T data and CMB measurements of DM relic density.
Plot-(a) presents the bounds in the $(\mX,\,\Lambda_S^{})$ plane
with $\Dm =2.8$\,keV and for the scalar-type
DM-electron interaction at the 68\%\,C.L.\ (red region)
and 95\%\,C.L.\ (pink region).
Plot-(b) depicts the bounds in the $(\mX,\,\Lambda_V^{})$ plane
with $\Dm =2.8$\,keV and for the vector-type
DM-electron interaction at 68\%\,C.L.\ (red region) and 95\%\,C.L.\ (pink region).
In each plot, the constraints by the DM relic density are
shown as blue and green curves, which are analyzed in Sec.\,\ref{sec:3}.
}
\label{fig:fit}
\label{fig:3}
\end{figure}

\vspace*{1mm}

We present our findings in Fig.\ref{fig:3}.
In plot-(a), we fit with the XENON1T data and
show the allowed parameter region in the $\mX\!-\!\Lambda_S^{}$
plane for the case of scalar-type DM-electron interaction
under $\Delta m=2.8\,\text{keV}$.
The black curve represents the best fit value,
and the red and pink bands give the allowed parameter space at
$68\%$\,C.L.\ and $95\%$\,C.L., respectively.
This parameter region is largely independent of the DM mass-splitting $\Delta m$,
as indicated by Fig.\ref{fig:2} which shows the recoil spectrum for different
$\Dm$ values under the same input ratio of $\,\over{\sigma}_e^{}/\mX$\,.
In parallel, we further present in plot-(b) the allowed parameter region in the
$\mX-\Lambda_V^{}$ plane for the case of vector-type DM-electron interaction
under $\Delta m=2.8\,\text{keV}$.
The $68\%$ and $95\%$ confidence limits of the XENON1T data
on the parameter space are marked by the red and pink colors, respectively.
In Fig.\,\ref{fig:3}, we also presented the constraints from the
CMB measurements on the DM relic density as the  blue and green curves which
we will derive and discuss in the next section.

\vspace*{1mm}

Finally, for the convenience of analysis, we present a compact formula for computing
the total number of DM signal events in the XENON1T experiment.
We define the following ratio which is mainly independent of the model-parameters
$(\Delta m,\,\,\mX ,\,\cut)$,
\beqa
\xi \,\equiv
\int\!\! \d\ER\,\eta(\ER)
\frac{1}{\bar{\sigma}_{\!e}^{}}
\frac{\,\d\langle\sigma_{\!Xe}^{}\vDM\rangle\,}
{{\rm d}\ER}\,,
\eeqa
where
$\eta(E)$ is the detector efficiency function. The only dependence
of $\,\xi\,$ on $(\Delta m,\,\mX)$ comes from the upper/lower limits $q_\pm^{}$
of the integration \eqref{eq:DSigDE}. We have shown in Fig.\,\ref{fig:2} that
varying $\Delta m$ will mainly shift the position of the recoil energy peak,
but has little effect on the height of the spectrum.
We have also checked numerically that the ratio $\,\xi\,$ only changes
by about $4\%$ when the DM mass $\,\mX\,$ varies within $(0.1\!-\!10)$\,GeV and
the mass-splitting varies within $\,\Delta m \!=\!(2-3)\,\text{keV}$.\,
As a benchmark point, we obtain $\,\xi_{0}^{}\!=\!1.62$\, for
$\Delta m\!=\!2.8\,$keV and $\mX\!=\!1\,$GeV.
Then, we derive the total number of excess events beyond the backgrounds:
\begin{align}
N_{\text{tot}}^{} & \simeq\,
\xi^{}_{0}\bar{\sigma}_{\!e}^{}\frac{\rho_{\text{DM}}^{}}{\,2\mX\,} N_T^{}\,\Delta t
\nn\\
& \simeq\, 50\!\times\!
\frac{\bar{\sigma}_{\!e}^{}}{\,8\!\times\!10^{-44}\text{cm}^{2}\,}
\frac{\rho_{\text{DM}}^{}}{\,0.3\,\text{GeV}\!/\text{cm}^{3}\,}
\frac{\,\text{GeV}\,}{\mX}\frac{\,N_{T}\,}
{\,4.2\!\times\!10^{27}\!/\text{ton}\,}
\frac{\Delta t}{\,0.65\,\text{ton}\,\text{yr}\,} \,.
\hspace*{8mm}
\end{align}
%

\vspace*{2mm}
\section{Constraints from DM Relic Abundance and Decays}
\label{sec:3}
\vspace*{1mm}

In this section, we compute the relic abundance for the inelastic DM and
analyze constraints on the DM parameter space
by following the conventional DM freeze-out mechanism\,\cite{Gondolo:1990dk,Srednicki:1988ce}.
We also derive constraint on the DM self-interactions.
Then, we show that the lifetime of the heavier DM component $X'$ can be much longer
than the age of the Universe.

\vspace*{1mm}

With the DM-lepton contact interactions,
we can compute the DM annihilation cross sections of
$XX',X'X',XX \!\!\to\!\ell^+\ell^-$
with $\,\ell \!=\! e,\mu,\tau$\,
for the scalar-type contact interactions \eqref{eq:O_SS}
and the vector-type contact interactions \eqref{eq:O_VL}-\eqref{eq:O_VR}.
For instance, the scalar-type operator $\mathcal{O}_S^{}$ contributes
to the annihilation cross section of
$XX'\!\!\to\! \ell^+\ell^-$ for $\,\mX,m_\ell^{}\!\gg\Dm$\,,
\begin{equation}
\sigma_{\text{ann}}^{S}
\,\simeq\, \frac{v^2}{\,8\pi\Lambda_S^4\,}\sqrt{\frac{s}{\,s\!-\!4m_{X}^{2}}\,}
\(\!1\!-\!\frac{4m_\ell^2}{s}\)^{\!\!\!3/2} ,
\end{equation}
while the vector-type operator ${\cal O}_{VL}^{}$ or ${\cal O}_{VR}^{}$
contributes to the annihilation cross section,
\begin{equation}
\sigma_\text{ann}^V \,\simeq\,
\frac{\,s\!-\!m_\ell^2\,}{\,24\pi\Lambda_V^{4}\,}
\sqrt{\(\! 1\!-\!\frac{\,4m_\ell^2\,}{s}\!\)
	  \(\!1\!-\!\frac{4m_X^2}{s}\!\)\,}
\,.
\end{equation}
The same formulas also hold for other initial states $XX$ and $X'X'$.
But for the vector-type operators
\eqref{eq:O_VL}-\eqref{eq:O_VR}, the initial state contains $XX'$ only.
Similar type of annihilation processes were considered in the literature\,\cite{Yu:2011by}.
Then, we further derive the thermal averaged annihilation cross section of
$\,XX'\!\to \ell^{+}\ell^{-}$
at the freeze-out temperature $T_{\!f}^{}$\,,
\begin{subequations}
\label{eq:sigma.v}
\begin{eqnarray}
\langle\sigma_\text{ann}^S v_{\text{DM}0}^{}\rangle &\!\!\simeq\!\!&
\frac{v^2}{\,4\pi\Lambda_S^{4}\,}
\(\!1-\frac{m_\ell^2}{m_X^2}\)^{\!\!3/2},
\label{eq:sigma.v-S}
\\[2mm]
\langle\sigma_\text{ann}^V \vDMM\rangle &\!\!\simeq\!\!&
\frac{m_X^2}{\,2\pi\, x_{\!f}^{}\Lambda_V^{4}\,}
\sqrt{1\!-\!\frac{m_\ell^2}{m_X^2}\,}\!
\(\!1\!-\!\frac{m_\ell^2}{\,4m_X^2\,} \!\) ,
\label{eq:sigma.v-V}
\end{eqnarray}
\end{subequations}
where $\vDMM$ is the DM velocity around the freeze-out epoch and
$\,x_{\!f}^{}\!= \mX /T_{\!f}^{}\!\simeq\! 18$\,.
In the above formulas we only keep the lowest order of $x_f^{-1}$.
For computing the DM relic density, we have determined the freeze-out temperature
and $x_{\!f}^{}$ numerically.
Following the analysis of the conventional freeze-out mechanism, we parameterize
$\sigma_\text{ann}\vDMM \simeq a_0^{} \!+\! a_1^{} x_f^{-1}$,
then the total relic density of $X$ and $X'$ is given by
\begin{equation}
\Omega_{\text{DM}}^{}h^{2} \,\simeq\,
2.1\!\times\!10^{9}\,\text{GeV}^{-1}\!
\(\!\frac{T_0^{}}{\,2.725\text{K}\,}\!\)^{\!\!3}
\!\frac{x_f^{}}{\,M_{\text{Pl}}^{}\sqrt{g_{*}(T_{\!f}^{})\,}
\left( a_0^{} \!+\! a_1^{} x_f^{-1}/2 \right) \,}\,,
\hspace*{15mm}
\end{equation}
where $M_{\text{Pl}}^{}$ denotes the reduced Planck mass and
$\,T_{\!f}^{}$\, is the freeze-out temperature.
We will determine $\,T_{\!f}^{}$\, and $x_{\!f}^{}$ numerically.
The temperature  $\,T_{0}^{}\simeq 2.725\,\text{K}$\, is
the current CMB temperature\,\cite{Fixsen:2009ug}
and $\,g_{*}^{}(T_{\!f}^{})$\, is the effective
degrees of freedom at the DM freeze-out.

\vspace*{1mm}

Using the CMB measurement of the current DM relic density
$\,\Omega_{\text{DM}}^{}h^{2}= 0.120\pm 0.001$ \cite{Aghanim:2018eyx},
we can derive strong constraints on the cutoff scales $\cut_S^{}$ and $\cut_V^{}$
for the scalar-type and vector-type DM interactions.
We present the bounds on $\cut_S^{}$ in Figs.\,\ref{fig:3}(a) and
the bounds on $\cut_V^{}$ in Figs.\,\ref{fig:3}(b), where in each plot the blue curve
depicts the bound from the annihilation channel $XX'\!\!\to\! e^{+}e^{-}$
and the green curve corresponds to the bound from all relevant annihilation channels
including the initial states $(XX',\, X'X',\, XX)$ and the final states
$(e^+e^-,\,\mu^+\mu^-,\,\tau^+\tau^-)$.
Note that for vector-type contact interaction, $XX'$ is the only possible initial state. To derive the green curve in each plot,
we have chosen a common cutoff scale for all the relevant
operators in Eq.\eqref{eq:O_SS} or Eqs.\eqref{eq:O_VL}\eqref{eq:O_VR}.
Fig.\,\ref{fig:3} shows that combining the DM relic density bound with the bound by
fitting the XENON1T anomaly, we can constrain the DM parameter space
$(\mX,\,\cut)$ into rather narrow regions.
For the case of scalar-type contact interactions, the relic density bounds in
Fig.\,\ref{fig:3}(a) are fairly flat, so the cutoff scale $\cut_S^{}$ is almost
fully fixed.  In summary, we obtain the following bounds on the DM mass and
the cutoff scale of the effective DM contact interactions,
%
\beqa
\hspace*{-14mm}
\text{Scalar-type:} \!\!&&\!\!
(\mX,\,\cut_S^{}) = (1.22\,\text{GeV},\,1.11\,\text{TeV}),~~
(\text{Best~Fit});
\nn\\
\hspace*{-14mm}
\!\!&&\!\!
(1.02\text{GeV},\,1.110\text{TeV})\!<\! (\mX,\,\cut_S^{}) \!<\! (1.90\text{GeV},\,1.113\text{TeV}),~~(\text{95\%\,C.L.}).
\label{eq:OS-bound}
\\
\hspace*{-14mm}
\text{Vector-type:} \!\!&&\!\!
(\mX,\,\cut_V^{}) = (2.95\,\text{GeV},\,74.8\,\text{GeV}),~~ (\text{Best~Fit});
\nn\\
\hspace*{-14mm}
\!\!&&\!\!
(2.48\text{GeV},\,68.2\text{GeV})\!<\! (\mX,\,\cut_V^{}) \!<\! (4.51\text{GeV},\,92.9\text{GeV}),~~(\text{95\%\,C.L.}).
\label{eq:OV-bound}
\eeqa
%
It shows that the bound on the vector-type cutoff scale $\cut_V^{}$ is much lower
than the bound on the scalar-type cutoff scale $\cut_S^{}$\,. The reason is because
the vector-type cross sections are much smaller than that of the scalar-type due to
the relative suppression factor $\,m_X^2/v^2=O(10^{-4})\,$ as shown in Eqs.\eqref{eq:sigma-S}-\eqref{eq:sigma-V}
and Eqs.\eqref{eq:sigma.v-S}-\eqref{eq:sigma.v-V},
where $\,v\simeq 174$\,GeV is the Higgs VEV.
We also note that the cutoff scale $\Lambda$ in the effective operator is
connected to the heavy mediator mass $M_{\text{md}}^{}$ via
$\,\Lambda = M_{\text{md}}^{}/\!\sqrt{\tilde{g}_X^{}\tilde{g}_\ell^{}}\,$,
where $\tilde{g}_X^{}$ denotes the mediator coupling to the DM and
$\tilde{g}_\ell^{}$ the mediator coupling to the leptons.
Thus, we can deduce $\,\Lambda\!\gg\! M_{\text{md}}^{}\,$
(when $\,\tilde{g}_X^{}\tilde{g}_\ell^{}\!\ll\! 1\,$),
or, $\,\Lambda\lesssim\! M_{\text{md}}^{}\,$
(when $\,\tilde{g}_X^{}\tilde{g}_\ell^{}\!\gtrsim\! 1\,$).

\vspace*{1mm}

The DM mass ranges in Eqs.\eqref{eq:OS-bound}-\eqref{eq:OV-bound}
are small enough, so the DM particles can be produced at the $e^+ e^-$ colliders\,\cite{LEP,DMcollider2,Essig:2013vha,Darme:2020ral}.
In the following, we summarize the current constraints by
the LEP and BaBar experiments from \cite{LEP,Essig:2013vha}%
\footnote{The bounds of \cite{LEP,Essig:2013vha} are given for the fermionic DM.
To derive the bounds for the scalar DM, we rescale the results of
\cite{LEP,Essig:2013vha} according to the cross section formulas of
the corresponding processes.}.
Note that the validity of the EFT requires
$\,M_{\text{md}}^2 \!\gg\! (2m_X^{})^2$\,.
Thus, for the simplicity of discussion,
we consider the case where the mediator mass is heavier than $10$\,GeV,
which is also about the center of mass energy of BaBar experiment.
For the scalar-type operator $\mathcal{O}_S^{}$\,, Eq.\eqref{eq:OS-bound}
gives a sizable cutoff scale $\cut_S^{}\!\simeq\! 1.1$\,TeV,
which is much above the current collider search limits of a few hundred GeV\,\cite{LEP,Essig:2013vha}.
The mediator in the UV completed model could receive additional constraints,
as we will discuss in Secs.\ref{sec:4.1}-\ref{sec:4.2}.
On the other hand, for the vector-type operator ${\cal O}_V^{}$,
Eq.\eqref{eq:OV-bound} gives a quite low cutoff scale
$\,\Lambda_V^{}\!\simeq\! (68\!-\!93)$\,GeV.
In the case of $\,\tilde{g}_X^{}\tilde{g}_\ell^{}\!\ll\! 1\,$,
we have $\,M_{\text{md}}^{}\!\ll\! \cut_V^{}$
and thus the mediator can be produced on-shell at the LEP.
The mediator contributes constructively to the cross section of
the lepton-pair production process
$\,e^+e^- \!\!\rightarrow\!\ell^+ \ell^-$.\,
This places a constraint
$\,\sqrt{4\pi s}/\tilde{g}_\ell^{} \!>\! 13.2$\,TeV
($95\%$\,C.L.)\footnote{%
Hereafter all the quoted experimental bounds are set at 95\%\,C.L.\
unless specified otherwise.}\,
for the mediator having universal coupling to all the leptons,
or, $\,\sqrt{4\pi s}/\tilde{g}_\ell^{} \!>\! 8.6$\,TeV
for the mediator having coupling to
electrons only\,\cite{LEP_l_scatt},
where $\sqrt{s}\simeq 200$\,GeV is the LEP collider energy.
This corresponds to
$\,\tilde{g}_\ell^{}\!\lesssim\! 0.054\,(0.082)$\,
for the case of universal coupling
(electron coupling only),
and in turn it imposes
$\,\tilde{g}^2_X \!\gg\! \tilde{g}^2_\ell $\,
for the low cut-off scale required by Eq.\eqref{eq:OV-bound}
and
$\,M_{\text{md}}^{} \!\gtrsim\! 10$\,GeV.\,
In this case, the mediator decays predominantly into
the DM particles\footnote{%
In this case, the LHC constraint via the decay process  $\,Z\!\to\! 4 \ell\,$ \cite{Sirunyan:2018nnz} becomes negligible.}
and is constrained by the mono-photon searches at LEP-2.
Since the width of the mediator scales as
$\,\Gamma_{\rm md}^{} \!\propto \tilde{g}_X^2 M_{\rm md}^{}$\,
and the mono-photon cross section scales as
$\,\sigma\propto \tilde{g}_\ell^2 \tilde{g}_X^2 M_{\rm md}^{} /(\Gamma_{\rm md} s)
\propto \tilde{g}_\ell^2/s$\,
for $\,s\!\gg\! M_{\rm md}^2$\,,
the mono-photon searches only constrain
$\,\tilde{g}_\ell^{}$\, and is independent of
$M_{\rm md}^{}$ in the mass range of interest.
Ref.\,\cite{Essig:2013vha} extracted such a bound
from the result of Ref.\,\cite{LEP} and sets
$\,\tilde{g}_\ell^{}\!\lesssim\! 0.023$\,,
or in terms of the cutoff scale, we can derive
\begin{equation}
\Lambda_V^{}\gtrsim\, 52\,{\rm GeV}\!\times\!
\(\!\!\frac{\,\sqrt{4\pi\,}\,}{\,\tilde{g}_X^{}}\!\)^{\!\!\!\frac{1}{2}}\!\!
\(\!\frac{M_{\rm md}^{}}{\,15\,{\rm GeV}\,}\!\) ,
\end{equation}
which is consistent with our bound \eqref{eq:OV-bound} within limited parameter space.
On the other hand, the constraint on the off-shell vector mediator
by the BaBar measurement sets a bound
$\,\Lambda_V^{}\!\gtrsim\! 30$\,GeV\,\cite{Essig:2013vha}\,
for $\,m_X^{}\!\lesssim\! 3$\,GeV.
The constraint becomes weaker for heavier DM due to the limited center
of mass energy. As the next generation B-factory,
the upcoming Belle-II experiment\,\cite{Abe:2010gxa}
is expected to further probe a much larger $\Lambda_V^{}$\,.
For the ideal case where the systematic errors are negligible,
 the Belle-II measurement can probe up to
$\,\Lambda_V^{}\!\lesssim 92$\,GeV\,\cite{Essig:2013vha},
and thus it could help to either confirm or rule out our EFT formulation of
the vector-type inelastic DM as a viable resolution to the XENON1T
anomaly.

\vspace*{1mm}

After the annihilation processes $\,XX',X'X',XX\!\!\to\!\ell^+\ell^-$
decouple, the total number of the DM particles $X$ and $X'$ is fixed.
But, the conversion between $X$ and $X'$ is still efficient due to the
scattering process $\,e^{\pm}X'\!\leftrightarrow\! e^{\pm}X$.\,
We may estimate the kinetic decoupling temperature of this process.
We note that for $\,T\!\lesssim\! m_e^{}$\,,
the scattering cross sections
$\bar{\sigma}_e^S$ and $\bar{\sigma}_e^V$
are already given by Eqs.\eqref{eq:sigma-S}-\eqref{eq:sigma-V}.
Thus, this decoupling happens when the following condition is realized,
\begin{equation}
\bar{\sigma}_e^{S,V} n_{e^\pm} \simeq\, H\,,
\end{equation}
where  $\,n_{e^\pm}\!\simeq\! 4\zeta(3)T^{3}/\pi^{2}$\,
is the number density of $e^\pm$, and
$\,H \!\simeq\!\sqrt{\!\frac{\,g_*^{}\pi^2}{90}}\frac{T^2}{\,M_\text{Pl}^{}}$\,
is the Hubble rate.
For the parameter space of $(\mX,\,\Lambda_{S}^{})$ or $(\mX,\,\Lambda_{V}^{})$
which realizes both the observed DM relic density and XENON1T signal excess,
we find that this conversion process freezes out at a temperature
$\,T\simeq 0.7\,\text{MeV} \!\gg\! \Dm$\,,
very close to the temperature when the $e^\pm$ density gets depleted.

\vspace*{1mm}

We note that the quartic interactions of the scalar DM contain a term
$\,\tilde{\lambda}X^2{X'}^2\,$ which induces the annihilation process
$\,X'X'\leftrightarrow XX\,$.
This converts the two types of DM particles into each other
and gives the following annihilation cross section,
\begin{equation}
\label{eq:X'X'XX}
\sigma[X'X'XX] \,\simeq\,
\frac{\tilde{\lambda}^{2}}{\,64\pi m_{X}^{2}\,} \,.
\end{equation}
After $e^\pm$ decouple from the dark sector, the temperature
of the dark matter drops quickly as $\,a(t)^{-2}$, with $\,a(t)\,$ being the
expansion scale factor of the Universe.
The DM temperature then falls below keV in
a very short time and an active annihilation $X'X'\!\to\! XX$
would deplete the $X'$ density.
Since the quartic scalar self-interaction $\,\tilde{\lambda}X^2{X'}^2\,$ is generally
independent of the DM-electron interactions, we may properly set the scalar coupling
$\,\tilde{\lambda}\,$ such that the annihilation $X'X'\!\to\! XX$ freezes out
before the electrons decouple.
The decoupling is realized by the condition
%
\beqa
\sigma[X'X'XX] \!\times\! v_{\text{DM}_0}^{}n_{\text{DM}}^{}
\,\simeq\, H \,,
\eeqa
where the DM has kinetic energy
\,$\fr{1}{2}\mX v_{\text{DM}_0}^{2}\!\!\approx\! \fr{3}{2}T$\,.
For the temperature $\,T\!\ll\!\mX$\,, we have
$\,n_{\text{DM}}^{} \!\approx\!
\frac{\pi^{2}}{15}\frac{\,\Omega_\text{DM}\,}{\Omega_\text{M}}
\frac{T_{\text{eq}}^{}}{\,\mX\,}T^3$,
where $T_{\text{eq}}^{}$ is the temperature at the matter-radiation equality;
and $\Omega_{\text{DM}}$ and $\,\Omega_\text{M}$ are the normalized present-day
DM density and total matter density, respectively.
By requiring the annihilation $\,X'X'\!\to\! XX$\, to freeze out before
$\,T\!\approx\!1\,\text{MeV}$, we deduce an upper bound on the DM self-coupling
$\,\tilde{\lambda}\lesssim 0.03\,$.
After $e^\pm$ decouple,
the ratio between the particle number densities of $X$ and $X'$
is frozen as $\,n_{X}^{}\!=n_{X'}^{}\!=\!\fr{1}{2}n_{\text{DM}}^{}$\,,
consistent with the setup throughout our formulation.

\vspace*{1mm}

Next, we further estimate the lifetime of the heavier dark matter component $X'$.
There are two possible decay channels, $\,X'\!\to\! X\gamma\,\gamma\,$
and  $\,X'\!\to\! X\nu\,\bar{\nu}\,$.
The decay rate is generally suppressed by the small DM mass-splitting
$\,\Delta m$\, which determines the energy scales of outgoing photons or neutrinos.
If $X$ and $X'$ couple to electrons through the contact interaction
${\cal O}_S^{}$ or ${\cal O}_{V\!R}^{}$,
then $X'$ decays dominantly into two photons $\,X'\!\to\! X\gamma\,\gamma\,$
through one-loop diagram with electron in the loop.
If $X$ and $X'$ couple to electrons through the contact interaction
${\cal O}_{V\!L}^{}$ instead, then $X'$ will decay dominantly via
invisible channels $\,X'\!\to\! X\nu\,\bar{\nu}\,$ at tree level.

\vspace*{1mm}

For the scalar-type interaction ${\cal O}_S^{}$,
we compute the one-loop diagram for $\,X'\!\to\! X\gamma\,\gamma\,$
and obtain the decay width:
\beqa
\Gamma_{X'}^{S}  \,\simeq\,
\frac{\,\alpha^{2}\,}{\,7560\pi^{5}\,}
\frac{\Delta m^7 v^2}{\,m_e^{2}m_X^2\Lambda_S^4\,}\,.
\eeqa
Here the electron triangle-loop has some similarity with the
SM Higgs decay into di-photons ($h\to\gamma\gamma$) via fermion
triangle-loop\,\cite{Ellis:1975ap}.
Taking the sample inputs of $\,\mX\approx 1$\,GeV and $\Lambda_S^{}\approx 1$\,TeV
for satisfying the constraints by the DM relic density and the XENON1T measurement,
we find the $X'$ lifetime as $\,\tau_{X'}^{}=O(10^{18})$\,yr,
which is 8 orders of magnitude longer than the age of the present Universe
($\sim\! 10^{10}$\,yr).
So it is far beyond any current constraints for the decaying dark matter.
Besides, we note that for the scalar-type interaction ${\cal O}_S^{}$,
the invisible decay channel $\,X'\!\to\! X\nu\,\bar{\nu}\,$
could occur via one-loop $W$-exchange triangle-loop. But its decay width is expected
to be highly suppressed by extra factors of $(m_{\nu}^2m_e^2)/M_W^4$
due to chirality-flips and thus fully negligible.

\vspace*{1mm}

For the vector-type interaction ${\cal O}_{V\!R}^{}$,
we compute its contribution to the decay width of $\,X'\!\to\! X\gamma\,\gamma\,$,
and obtain to the leading order of $\,\Delta m$\,,
\begin{equation}
\Gamma_{X'}^{V\!R}  \,\simeq\,
\frac{\alpha^{2}}{\,7560\pi^5\,}
\frac{\Delta m^{9}}{\,m_e^4 \Lambda_V^4\,}\, .
\end{equation}
This loop result is consistent with that of \cite{Jackson:2013pjq} when comparable.
From the above formula, we deduce the $X'$ lifetime
$\,\tau_{X'}^{V\!R}=O(10^{23})$yr
for $\,\Lambda_V^{}\!\sim\!100$\,GeV.
This is again far beyond the age of the Universe.
For the vector-type interaction ${\cal O}_{V\!R}^{}$,
we may further consider the invisible decay channel $\,X'\!\to\! X\nu\,\bar{\nu}\,$
via $W$-exchange triangle-loop. But we find that the $X'$ decay width
is highly suppressed by an extra chirality-flip factor of
$\,m_e^4/M_W^4\,$, so it is fully negligible.

\vspace*{1mm}

Then, for the vector-type interaction ${\cal O}_{V\!L}^{}$,
we see that $X'$ will decay predominantly via the invisible channel
$\,X'\!\to\! X\nu\,\bar{\nu}\,$ at tree level. We can derive its decay rate,
\begin{equation}
\Gamma_{X'}^{V\!L} \,=\,
\frac{1}{\,120\pi^{3}\,}
\frac{\Delta m^5}{\,\Lambda_{V\!L}^4\,} \,.
\end{equation}
By requiring the $X'$ decay lifetime larger than the age of the present Universe
(about $1.38\!\times\!10^{10}$yr) and inputting the fitted range of DM mass-splitting
$\,2.1\,\text{keV}\!<\!\Dm\! <3.3\,\text{keV}\,$ (95\%\,C.L.)
from Fig.\,\ref{fig:1}(b),
we derive the lower bound on the cutoff scale
$\,\cut_V^{}\!>\!(291\!-512)$GeV.
Comparing this with the allowed range given in Fig.\,\ref{fig:3}(b), we find that
this is excluded by both the DM relic density measurement and the current XENON1T data.
Hence,  the vector-type interaction ${\cal O}_{V\!L}^{}$ cannot provide a viable
inelastic DM resolution.

\vspace*{1mm}

Finally, we note that the effective DM-electron interactions
can also induce new decay channels of the SM gauge bosons $W/Z$ and
the Higgs boson $h^0$
with $X$ and $X'$ in the decay products. But, such decays
are realized by either attaching the effective DM-electron quartic vertex to an electron
loop or to an out-going electron line in the Feynman diagram. In both cases,
the corresponding decay width and branching fraction are suppressed by an extra factor of
$\,(1/16\pi^2)^{2}\!\lesssim\! 10^{-4}$,\,
either from the loop factor or from the phase space of two
additional DM particles in the final state.
Hence such effects are far below the current experimental sensitivity\,\cite{PDG}.

\vspace*{2mm}
\section{UV Completion for Effective DM-Lepton Interactions}
\label{sec:4}
\label{sec:UV-completions-of}
\vspace*{1mm}

In this section, we study possible UV completions for the effective
DM-lepton interactions
${\cal O}_S^{}$ and ${\cal O}_{V\!R}^{}$ as illustration.
In the first model, the effective DM-lepton interactions are mediated by
an extra heavy Higgs doublet.
In the second model, the interactions are mediated by extra vector-like heavy leptons.
In the third model, the interactions are mediated by a new gauge boson that couples to
the leptons and DM. We also note that whenever a light singlet scalar DM is coupled to
the Higgs sector, there are Higgs portal terms such as
$\lambda_{X\!H}^{}X^{2}|H|^{2}$, $\lambda_{X'\!H}^{}X'^{2}|H|^{2}$ and
$\lambda_{X\!X'\!H}^{}XX'|H|^{2}$.
These interactions will induce invisible decays of the SM Higgs boson,
so their couplings are constrained by the Higgs boson measurements at the LHC, $\lambda_{X\!H}^{},\lambda_{X'\!H}^{},\lambda_{X\!X'\!H}^{}\!\lesssim 10^{-2}$
\cite{Escudero:2016gzx}.

\subsection{Mediation by Second Higgs Doublet}
\label{sec:4.1}

In this model, we couple the real scalar DM fields $X$ and $X'$ to a
two-Higgs-doublet model (2HDM)\,\cite{2HDM}.
The relevant terms in the Lagrangian are
\begin{equation}
\label{eq:2HDM-XXH1H2}
{\cal L} \,\supset \, y'_{j}\bar{L}_{j}^{}H_{2}^{}\ell_{jR}^{}
+\lambda_{12}' XX'H_{2}^{\dagger}H_{1}^{\vphantom{\dagger}}+ \text{h.c.},
\end{equation}
where $H_1^{}$ and $H_2^{}$ are two Higgs doublets, and
lepton $\,\ell_j^{}\!=\!e,\mu,\tau$\,.
For convenience, we may arrange the Higgs potential such that $H_{1}^{}$
is a SM-like Higgs doublet with the full VEV
$\langle H_{1}^{}\rangle\!=\!(0,v)^{T}$,
and $H_{2}^{}$ is a heavy Higgs doublet with vanishing VEV
$\langle H_2^{}\rangle\!=\!(0,0)^{T}$.\,
This means that $H_2^{}$ is irrelevant to fermion mass-generation, so
its Yukawa couplings such as $y_{\ell}'$ can be very different from
the leptonic Yukawa coupling $\,y_{\ell}^{}\!=\! m_\ell^{}/v$\, in the SM.
As before, we assign $X$ and $X'$ to be odd under an exact
$\mathbb{Z}_2^{}$ symmetry which ensures the DM stability,
while all other fields are $\mathbb{Z}_2^{}$ even.
For $H_{2}^{}$ being a heavy Higgs doublet,
we can integrate it out and induce the following
effective operator at low energies,
\begin{equation}
{\cal O} \,=\,
\frac{\,y'_{j}\lambda_{12}'\,}{\,M_{H_2^{}}^{2}\,}
\bar{L}_{j}^{}H_{1}^{}\ell_{jR}^{}XX'+ \text{h.c.}
\end{equation}
This just gives the effective dimension-6 operator
${\cal O}_S^{}$ in Eq.\eqref{eq:O_S} with the cutoff scale
$\,\Lambda_{S}^{}=M_{H_{2}^{}}^{}/\!\sqrt{|y'_{j}\lambda_{12}'|}$\,.
The quartic interaction in Eq.\eqref{eq:2HDM-XXH1H2}
can also induce a contribution to the DM self-interaction term
$\,\delta\tilde\lambda \,X^2X'^2$\, with
$\,\delta\tilde\lambda = \lambda_{12}^{\prime\,2}\,v^2/M_{H_2^{}}^2$.
To generate the observed relic density and explain the XENON1T excess,
we require $\,\Lambda_S^{}\simeq 1.1$\,TeV as in Eq.\eqref{eq:OS-bound}.
For $\,y'_{e}\lambda_{12}'=O(1)$\,,
this requires the mass of the second Higgs doublet to be
$\,M_{H_2^{}}^{}\!=O(1)$\,TeV.

\subsection{Mediation by Vector-like Heavy Leptons}
\label{sec:4.2}

In this subsection, we consider the second model
where the effective interaction is mediated by a new
generation of vector-like heavy leptons.
The setup has some similarity to the
lepton portal DM model\,\cite{Bai:2014osa}\cite{Ge:2020tdh},
but it contains both vector-like fermion singlets and doublets as the mediators,
which have mixings induced by Higgs VEV via Yukawa-type interactions.
If coupled to the muon, such extra mixed vector-like leptons can also be a
potential resolution to the muon $g_\mu^{}\!-2$ \cite{Kawamura:2020qxo}.
This model contains the following new terms beyond the SM Lagrangian,
\beqa
\Delta{\cal L}  \,=\,
\left[ y_{X}^{}\bar{L}_{j}^{}\Psi X
+y_{X'}^{}\bar{f}\ell_{jR}^{}X'
+y'\bar{\Psi}Hf + \text{h.c.}\right]
+M_{\!f}^{}\bar{f}f
+M_{\Psi}^{}\bar{\Psi}\Psi \,,
\label{eq:mixed_lepton_portal}
\eeqa
where the Dirac fermion
$\Psi$ is an $SU(2)_{L}^{}$ doublet with hypercharge
$\,Y_{\Psi}^{}\!=\!-\frac{1}{2}$,\,
and Dirac fermion $f$ is a weak singlet with hypercharge
$\,Y_{f}^{}\!=\!-1$.\, Both the fermions $\Psi$ and $f$ are $\mathbb{Z}_2^{}$ odd,
just like the DM $X$ and $X'$.
We also set a small coupling for the terms $y'_X\bar{L}_{j}^{}\Psi X'$ and $y'_{X'}\bar{f}\ell_{jR}^{}X$.
If $y'_X$ and $y'_{X'}$ are as large as $y_{X}^{}$ and $y_{X'}^{}\,$,
the electron anomalous magnetic moment
$\,g_e^{}-2$\, would receive an unacceptably large correction.
In this case, the annihilation cross section
$\,\sigma_{0}^{}\!\sim\! 10^{-9}\,\text{GeV}^{-2}$
is required for the DM relic density and can be related to
$\,\Delta(g_e^{}\!-\!2)\!\sim\!
\frac{m_{e}^{}}{16\pi^{2}}\sqrt{2\pi\sigma_{0}^{}} \!\sim\! 10^{-10}$
\cite{Kawamura:2020qxo}.
Thus, we suppress the couplings for
$\,y'_X\bar{L}_{j}^{}\Psi X'\,$ and
\,$y'_{X'}\bar{f}\ell_{jR}^{}X'$
in this model setup.
Although these two terms could be generated
by one-loop diagrams in connection to the leptons,
they are suppressed by the small SM
lepton Yukawa couplings $y_{\ell j}^{}$.\,
To see this explicitly, we note that in the limit of setting the couplings
$\,y_{\ell j}^{},y'_X, y'_{X'}\!=0$\,, the Lagrangian
\eqref{eq:mixed_lepton_portal} is invariant
under a discrete $\mathbb{Z}_2'$ symmetry:
$\Psi\!\rightarrow\! -\Psi$,
$\,X\!\rightarrow\! -X$, $\ell_{jR}^{}\!\rightarrow\! -\ell_{jR}^{}$,
and $\,f\!\rightarrow\! -f$\,.
This symmetry is broken by the SM lepton Yukawa couplings $y_{\ell j}^{}$.
Hence, the loop-generated couplings $y'_X$ and $y'_{X'}$
are proportional to $y_{\ell j}^{}$\,.

\vspace*{1mm}

Integrating out the heavy vector-like fermions $\Psi$ and $f$\,,
we obtain the following gauge-invariant dimension-6 effective operator,
\beqa
{\cal O} \,=\,
\frac{\,y_{X}^{}y_{X'}^{}y'}{\,M_{\Psi}^{}M_{\!f}^{}\,}
\bar{L}_{j}^{}H \ell_{jR}^{} XX' + \text{h.c.}
\eeqa
This is just the scalar-type operator ${\cal O}_S^{}$ given in
Eq.\eqref{eq:O_S}, with the cutoff scale
$\,\Lambda_S^{} \!=\!
\left[ M_{\Psi}^{}M_{\!f}^{}/(y_{X}^{}y_{X'}^{}y')\right]^{1/2}$.
From our analysis in Section\,\ref{sec:3},
we find $\,\Lambda_S^{}\!\approx\! 1.1$\,TeV in order
to realize the observed relic density and explain the XENON1T signal excess.
For $\,y_{X}^{}y_{X'}^{}y' \!\lesssim\! 1$\, and
$\,M_{\Psi}^{}\!\approx\! M_{\!f}^{}$,
this suggests  $M_{\Psi}^{},M_{\!f}^{}\!\gtrsim\! 1.1\,$TeV,
which is above the current collider limit of
$900$\,GeV on the vector-like leptons\,\cite{Kawamura:2020qxo}\cite{LHC-bound}.

\subsection{UV Completion for Vector-type Contact Operator} 
\label{sec:4.3}

In this subsection, as an illustration
we provide a UV completion for the vector-type
effective operator ${\cal O}_{V\!R}^{}$ in Eq.\eqref{eq:O_VR}.
We introduce an extra dark $U(1)_X^{}$ gauge group with gauge boson
$A'_\mu$ and two complex scalar singlets $S$ and $S'$.
The electroweak symmetry is spontaneously broken by the SM-like
Higgs doublet $H$ with VEV
$\langle H \rangle\!=\!(0,v_h^{})^{T}$.
The dark $U(1)_X^{}$ gauge group is broken at the TeV scale
by the singlet scalar field $S$ with VEV
$\,\left<S\right>\!=v_S^{}/\!\sqrt{2}$\,.
We present the anomaly-free particle content and group assignments
in Table\,\ref{tab:OVR_UV}.

\vspace*{1mm}

We can write down the relevant new Lagrangian terms as follows,
%
\begin{eqnarray}
\Delta {\cal L} &\!\!=\!\!&
\bar{\ell}_{R}^{}\ii\slashed{D} \ell_{R}^{}
+ |D^\mu \widehat{X}|^2
- m_{\widehat{X}}^2|\widehat{X}|^2
- \lambda_{\widehat{X}\!S} |S|^2 |\widehat{X}|^2
- \lambda_{\widehat{X}\!H} |H|^2 |\widehat{X}|^2
\nonumber\\[1mm]
&&
+|D^\mu S|^2 \!+\! |D^\mu S'|^2
\!+\mu_S^2|S|^2 \!-\!\lambda_S^{}|S|^4
\!-\!M_{S'}^2|S'|^2 \!-\!\lambda_{S'}^{}|S'|^4
\nn\\[1mm]
&&
-\lambda_{SS'} |S|^2|S'|^2
+ \!\left(\tilde\lambda_{SS'}^{} S'S^3
\!+\! \tilde{\lambda}_{\widehat{X}\!S'}^{}
S^{\prime\,2}\widehat{X}^2 \!+ \text{h.c.}
\!\)
-\(y_S^{} \nu_R^T S \nu_R^{} + \text{h.c.}\)
+\, \cdots ,
\hspace*{13mm}
\label{eq:DL}
\end{eqnarray}
where we have suppressed the fermion family indices for
simplicity of notations. In the above, we consider that all the scalar
couplings respect CP symmetry and are thus real.
We note that the VEV of the singlet $S$ can generate
TeV-scale Majorana masses for the right-handed neutrinos,
$\,M_R^{}\!=\!\sqrt{2}\,y_S^{}v_S^{}=O(0.5)$TeV,
for the sample inputs of the scalar VEV
$\,v_S^{}\!=\!O(100)\text{GeV}$\, and the Yukawa coupling
$\,y_S^{}\!=\!O(3)$.
Then, we can generate the light neutrino masses through
a TeV scale seesaw mechanism,
\begin{equation}
\label{eq:nu-seesaw}
m_\nu^{}=\frac{\,y_\nu^2\,v_h^2~}{\,M_R^{}\,}\,.
\end{equation}
We consider that the $\nu_R^{}$ Yukawa couplings
with the Higgs and lepton doublets to be
$\,y_\nu^{}\!=O(y_e^{})$, where
$\,y_e^{}\simeq 3\!\times\!\!10^{-6}$
denotes the electron Yukawa coupling
in the standard model (SM).
Thus, we find that Eq.\eqref{eq:nu-seesaw} provides
the light neutrino masses $\,m_\nu^{}\!=\!O(0.1)\,$eV,
which are consistent with the neutrino oscillation data.

\begin{table}[t]
	\begin{center}
	\renewcommand{\arraystretch}{1.5} 
	\begin{tabular}{c||ccc|ccc|c|c|c|c}
\hline\hline
Group  & $Q_{jL}^{}$ & $u_{jR}^{}$ & $d_{jR}^{}$ & $L_j^{}$ & $\ell_{jR}^{}$
		& $\nu_R^{}$ & $H$ & $S$
		& $S'$ &
		$\widehat{X}$ \\
\hline\hline
$SU(2)_L^{}$ & $\bf 2$ & $\bf 1$ & $\bf 1$ & $\bf 2$ & $\bf 1$
& $\bf 1$ & $\bf 2$ & $\bf 1$ &  $\bf 1$ & $\bf 1$ \\
\hline
$U(1)_Y^{}$ & $\frac{1}{6}$ & $\frac{2}{3}$ & $-\frac{1}{3} $
		& $-\frac{1}{2}$ & $-1$ & $0$ & $\frac{1}{2}$ & $0$
		& $0$ & $0$ \\
\hline
$U(1)_X^{}$ & $0$ & $\frac{1}{2}$ & $-\frac{1}{2}$ & $0$ & $-\frac{1}{2}$
& $\frac{1}{2}$ & $\frac{1}{2}$ & $-1$ &$3$ & $-3$ \\
\hline
$\mathbb{Z}_2^{}$ & $+$ & $+$ & $+$ & $+$ & $+$ & $+$ & $+$ & $+$ & $+$ & $-$ \\
\hline\hline
\end{tabular}
\renewcommand{\arraystretch}{1} 
\end{center}
\vspace*{-3mm}
\caption{\small Particle content and group assignments
of an UV completion model for the vector-type
contact operator ${\cal O}_{V\!R}^{}$\,. Here $Q_{L_j}^{}$ and $L_j^{}$ denote
the left-handed weak doublet of the SM quarks and leptons, respectively,
and the subscript $j$ is the fermion family index of the SM.}
\label{tab:OVR_UV}
\label{tab:1}
\vspace*{2mm}
\end{table}

\vspace*{1mm}

For the singlet $S'$ having a large positive mass
$M_{S'}^{}\!=\!O(10)$TeV, we can minimize the scalar potential
of Eq.\eqref{eq:DL} for $S'$ and realize a naturally small VEV
$\,\left<S'\right>\equiv v_{S'}^{}/\sqrt{2}$\, as follows,
\beqa
v_{S'}^{} \,\simeq\,
\frac{~\tilde\lambda_{SS'}^{}v_S^3~}{2M_{S'}^2}\,,
\label{eq:vS'}
\eeqa
where we have ignored the tiny ratios in the denominator (relative to the
large mass term $M_{S'}^2$),
$\,\lambda_{SS'}^{}v_S^2/M_{S'}^2\!\ll\! 1\,$ and
$\,\lambda_{S'}^{}v_{S'}^2/M_{S'}^2\!\simeq\!
\fr{1}{4}\lambda_{S'}^{}\tilde\lambda_{SS'}^2(v_{S}^{}/M_{S'}^{}\!)^6\!\lll\! 1\,$,
for $\,\lambda_{SS'}^{},\tilde\lambda_{SS'}^{}\!=\!O(10^{-1})\,$ and
$\,\lambda_{S'}^{}\!\lesssim\! O(1)\,$.
From Eq.\eqref{eq:vS'} and taking the sample inputs
$M_{S'}^{}\!=\!O(3)$TeV, $\,v_S^{}\!=\!O(100)$GeV and
$\,\tilde\lambda_{SS'}^{}=O(10^{-1})$,\,
we deduce a small $S'$ VEV,
$\,v_{S'}^{}\!=O(\text{MeV})$.

\vspace*{1mm}

The complex singlet
$\,\widehat{X}\!=\!(X\!+\!\ii X')/\!\sqrt{2}$\,
contains the DM components $X$ and $X'$, which are stabilized
by the  $\mathbb{Z}_2$ parity defined in Table\,\ref{tab:1}.
The $(X,\,X')$ mass-splitting is generated by a quartic term
$\,\tilde\lambda_{\widehat{X}\!S'}^{}{S'}^2\widehat{X}^2\!+\text{h.c.}$,
as shown in Eq.\eqref{eq:DL}.
For the setup in Eq.\eqref{eq:DL}, we derive the mass-splitting of
$(X,\,X')$ after spontaneous symmetry breaking,
\begin{subequations}
\begin{align}
m_X^{2} & =\, m_{\widehat{X}}^2 +
	\!\(\!\frac{\,\lambda_{\widehat{X}\!S}v_S^2\,}{2}
	+\lambda_{\widehat{X}H}v_{H}^2
	- \tilde\lambda_{\widehat{X}\!S'}v_{S'}^2\!\)\!
	+\cdots
\,,\\
m_{X'}^{2} & =\, m_{\widehat{X}}^2 +
	\!\(\!\frac{\,\lambda_{\widehat{X}\!S}v_S^2\,}{2}
	+\lambda_{\widehat{X}H}v_{H}^2
	+ \tilde\lambda_{\widehat{X}\!S'}v_{S'}^2\!\)\!
	+\cdots\,.
\end{align}	
\label{eq:DM-masses}
\end{subequations}
\hspace*{-3mm}
We can realize $\,m_X^2,m_{X'}^2=O(m_{\widehat{X}}^2)$\,
by setting the mixed quartic couplings
$\lambda_{\widehat{X}\!S}^{},\lambda_{\widehat{X}\!H}^{}\!\!=\!0$\,
at tree level.
From Eq.\eqref{eq:DM-masses}, we derive the DM mass-splitting,
\begin{equation}
\frac{\,m_{X'}^{}\!-\!m_{X}^{}\,}{m_X^{}}
\,\simeq\,
\frac{\,\tilde\lambda_{\widehat{X}\!S'}^{}v_{S'}^2\,}{m_X^2}\,.
\label{eq:mX'-mX}
\end{equation}
Below Eq.\eqref{eq:vS'}, we obtained the sample value of the
singlet scalar VEV $\,v_{S'}^{}\!\!=\!O(\text{MeV})$.
With the inputs
$\,m_X^{}\!=\!O(\text{GeV})$ and
$\tilde\lambda_{\widehat{X}\!S'}^{}\!\!=\!O(1)$,
we can further derive the desired DM mass-splitting
$\,\Delta m=O(\text{keV})\,$
from Eq.\eqref{eq:mX'-mX}.

The $U(1)_X^{}$ gauge boson $A'_\mu$ acquires a mass
after spontaneous gauge symmetry breaking.
For the case of $\,g_X^{}\!\ll g,g'$, we have
\begin{equation}
M_{\!A'}^{} \simeq\, g_X^{}\!\(\frac{1}{2}v_h^2\!+\!v_S^2\)^{\!\!\frac{1}{2}} ,
\end{equation}
where $g$, $g'$ and $g_X^{}$ are the gauge couplings
of $SU(2)_L^{}$, $U(1)_Y^{}$ and $U(1)_X^{}$, respectively.
Integrating out the massive gauge field $A'_\mu$\,,
we derive the dimension-6 effective operator
${\cal O}_{V\!R}^{}$ in Eq.\eqref{eq:O_VR} with a cut-off scale:
\begin{equation}
{\Lambda_{V\!R}^{}} =
\sqrt{\frac{2}{3}}
\frac{\,M_{\!A'}^{}}{g_X^{}}\, .
\end{equation}
We note that the DM particle in this model not only induces
the effective operator ${\cal O}_{V\!R}^{}$,
but also couples to right-handed quarks.
This opens up new DM annihilation channels in the early universe.
Thus, the constraint set by the DM relic density in Section\,\ref{sec:3}
is inapplicable to this model while the constraint
by the XENON1T data in Section\,\ref{sec:2.2} remains valid.
Since the Higgs doublet is charged under $U(1)_X^{}$, the model
is further constrained by electroweak precision tests.
We will explore the experimental tests of this model
and related phenomenology elsewhere.

\vspace*{2mm}
\section{Conclusions}
\label{sec:5}
\label{sec:Conclusion}
\vspace*{1mm}

Probing the dark matter (DM) signals via electron recoil provides an important means
for direct detection of light DM particles.
In this work, we explored an attractive resolution of the newly
reported XENON1T anomaly via exothermic inelastic scattering between
the DM particles and electrons. In this scenario, the dark matter sector
contains two components $X$ and $X'$ with a small mass-splitting
$\,\Delta m = m_{X'}^{}\!-\!m_{X}^{}$\,
close to the recoil energy of the excess events. The inelastic scattering of
the heavy component $X'$ with electrons de-excites it to the lighter state $X$,\,
releasing the energy to the recoiled electrons.

\vspace*{1mm}

In Section\,\ref{sec:2}, we presented an effective field theory (EFT) approach to
inelastic DM signals for the Xenon electron recoil detection.
For relatively heavy mediator, we formulated the DM-lepton interactions by
gauge-invariant effective contact operators of dimension-6 which contains
two DM fields $(X,\,X')$ and two leptons,
as given in Eqs.\eqref{eq:O6}-\eqref{eq:O_SS}.
Then, we computed the electron recoil energy spectrum
and fitted the XENON1T data. We found that the
DM mass-splitting falls into the range
$\,\Dm \!=\!(2.1\!-\!3.3)\,$keV at 95\%\,C.L., with the best fit
$\,\Dm \!=\!2.8\,$keV, which is shown in Fig.\,\ref{fig:1}
and Fig.\,\ref{fig:2}.

\vspace*{1mm}

In Section\,\ref{sec:3}, we analyzed the relic abundance for the inelastic DM.
The DM particles were in kinetic and chemical equilibrium in the early Universe.
The DM relic abundance is determined by the conventional freeze-out mechanism.
The conversion between the heavier and lighter DM states was maintained by
their scattering with $e^\pm$ in the plasma. The conversion became inefficient at
$\,T\!\approx\! 1$\,MeV and the proportion of the two DM components
was frozen at $\,n_{X}^{}\!\simeq n_{X'}^{}$.\,
We derived constraint on the DM self-interactions $\,\tilde{\lambda}X^2{X'}^2\,$
to ensure that the DM annihilation $X'X'\!\to\! XX$ froze out
before the $e^\pm$ decoupled.
We also found that the decay of the heavier component $X'$
is severely suppressed by the small DM mass-splitting $\,\Dm\,$,
so its lifetime is much longer than the age of the Universe.
This means that the DM inelastic scattering
$\,X'\,e^-\!\!\to\! X\, e^{-}\,$ still happens in the Universe today.
We further identified the viable parameter space to realize the observed
DM relic abundance and the XENON1T recoil energy spectrum,
as shown in Fig.\,\ref{fig:3} and Eqs.\eqref{eq:OS-bound}-\eqref{eq:OV-bound}.

\vspace*{1mm}

Finally, in Section\,\ref{sec:4} we presented three plausible UV completions
for the effective operators \eqref{eq:O6}-\eqref{eq:O_SS}. The first model
is given in Section\,\ref{sec:4.1}, which
is a 2HDM extension with an extra heavy Higgs doublet as the mediator to
induce the scalar-type DM-lepton interactions.
The second model is shown in Section\,\ref{sec:4.2}.
It contains extra vector-like heavy leptons as mediators to generate scalar-type
DM-lepton interactions.
For illustration, we presented the third model in Section\,\ref{sec:4.3},
in which the DM-lepton interactions are mediated by the new gauge boson $A_\mu'$
of a dark $U(1)_X^{}$ gauge group. This gauge group is spontaneously broken
at the weak scale
and a weak scale seesaw mechanism is realized for mass-generation of light
neutrinos. At low energies, the dark gauge boson exchange can induce the vector-type
DM-lepton interactions.

\vspace*{1mm}

We stress that our generic EFT approach
in Sections\,\ref{sec:2}-\ref{sec:3} has provided a valuable means for
studying the inelastic DM and its implications for the Xenon electron recoil detection.
With this approach, we identified new viable parameter
space of the inelastic DM as in Figs.\,\ref{fig:1}-\ref{fig:3}, and
realized the inelastic DM via attractive UV-completion models in
Section\,\ref{sec:4}. These will be further tested via the electron recoil measurements
by the next-generation DM detectors,
including the upcoming experiments of PandaX-4T\,\cite{PandaX4T}, LZ\,\cite{LZ},
and XENONnT\,\cite{XENONnT}.

\vspace*{5mm}
\appendix

\noindent	

\section{Independent Operators for DM-Lepton Interactions}
\label{app:A}

In this Appendix, we demonstrate that the effective operators in Eq.\eqref{eq:O6}
are the general gauge-invariant dimension-6 operators which are relevant
for studying the inelastic DM-electron scattering.
In the following,
we focus on the operators including the DM bilinear fields of $X$ and $X'$,
\begin{subequations}
\beqa
{\cal O}_S^{} &\!\!=\!&
(\bar{L}H \ell_{R}^{})(XX') + \text{h.c.} \,,
\label{eq:LHeXX}\\
{\cal O}_{V\!L}^{} &\!\!=\!&
(\bar{L}\gamma^{\mu}L)
\!\left(X'\partial_{\mu}^{}X-X\partial_{\mu}^{}X'\right),
\label{eq:LLXX-}\\
{\cal O}_{V\!R}^{} &\!\!=\!&
(\bar{\ell}_{R}^{}\gamma^{\mu}\ell_{R}^{})
\!\left(X'\partial_{\mu}^{}X-X\partial_{\mu}^{}X'\right).
\label{eq:eeXX-}
\eeqa
\end{subequations}
We show that the other relevant operators can be reexpressed
in terms of this set of operators.

\vspace*{1mm}

In general, we may also write down the following dimension-6 operators:
\begin{subequations}
\begin{align}
& (\bar{L}\ii\slashed{D}L)(XX') \,,
\label{eq:LDLXX}\\
& (\bar{\ell}_{R}^{} \ii\slashed{D} \ell_{R}^{})(XX') \,,
\label{eq:eDeXX}\\
& (\bar{\ell}_{R}^{}\gamma^{\mu}\ell_{R}^{})
\!\left(X'\partial_{\mu}^{}X\!+\!X\partial_{\mu}^{}X'\right) \!,
\label{eq:eeXX+}
\\
& (\bar{L}\gamma^{\mu}L)
\!\left(X'\partial_{\mu}^{}X\!+\!X\partial_{\mu}^{}X'\right)\!,
\label{eq:LLXX+}
\\
&
B^{\mu\nu}\partial_{\mu}X\partial_{\nu}X' ,
\label{eq:BXX}
\end{align}
\end{subequations}
where $D^{\mu}$ is the covariant derivative and
$B_{\mu\nu}$ is the field strength tensor of the SM hypercharge gauge group $U(1)_Y^{}$.
For the operators \eqref{eq:LDLXX}
and \eqref{eq:eDeXX}, they can be converted to the form of Eq.\eqref{eq:LHeXX}
with additional suppression by the small leptonic Yukawa
coupling $\,y_{\ell}^{}\,$ after applying the equations of motions (EOM),
\beqs
\begin{align}
{\rm i}\slashed{D}L & \,=\, y_{\ell}^{}H
\ell_{R}^{}+\cdots\,,
\label{eq:eR_EOM}\\
{\rm i}\slashed{D}\ell_{R}^{} & \,=\,
y_{\ell}^{}H^{\dagger}L+\cdots \,.
\end{align}
\eeqs
Hence, the contributions of the operators \eqref{eq:LDLXX}-\eqref{eq:eDeXX}
are negligible for the present study.

\vspace*{1mm}

For Eq.\eqref{eq:eeXX+}, we note that up to integration by part,
a total derivative term in the Lagrangian gives vanishing contribution
and leads to the following:
\begin{equation}
0\,=\,
\partial_{\mu}^{}
(\bar{\ell}_{R}^{}\gamma^{\mu}\ell_{R}^{} XX')
=\partial_{\mu}^{}
(\bar{\ell}_{R}^{}\gamma^{\mu}\ell_{R}) (XX')
+(\bar{\ell}_{R}^{}\gamma^{\mu}\ell_{R}^{})
(X'\partial_{\mu}^{}X \!+\! X\partial_{\mu}^{}X') \,.
\end{equation}
If we set the small leptonic Yukawa coupling $\,y_\ell^{}\!=\!0\,$,
then the lepton chirality is conserved at tree level
and thus $\,\partial_{\mu}^{}(\bar{\ell}_{R}^{}\gamma^{\mu}\ell_{R}^{})\!=\!0$\,.
Hence, including the leptonic Yukawa couplings only leads to a term suppressed
by $\,y_{\ell}^{}$\,.
To see this explicitly, we apply the EOM \eqref{eq:eR_EOM} and obtain $\,\partial_{\mu}^{}(\bar{\ell}_{R}^{}\gamma^{\mu}\ell_{R}^{})
= y_{\ell}^{}\bar{\ell}_{R}^{}H^{\dagger}L+\text{h.c.}+\cdots$.
Thus, we arrive at
\begin{equation}
(\bar{\ell}_{R}^{}\gamma^{\mu}\ell_{R}^{})
(X'\partial_{\mu}^{}X \!+\! X\partial_{\mu}^{}X' )
\,=\, y_{\ell}^{}(\bar{\ell}_{R}^{} H^{\dagger}L)(XX')+ \text{h.c.} + \cdots ,
\end{equation}
which again reduces to the form of Eq.\eqref{eq:LHeXX},
but suppressed by the small leptonic Yukawa coupling $y_{\ell}^{}$\,.
The exactly same reasoning holds for the operator \eqref{eq:LLXX+}.

\vspace*{1mm}

For Eq.\eqref{eq:BXX}, the following total derivative term gives
vanishing contribution up to integration by part and leads to:
\beqa
0 \,=\, \partial_{\mu}\!\left[B^{\mu\nu}(X\partial_{\nu}X')\right]
=\(\partial_{\mu}B^{\mu\nu}\)(X\partial_{\nu}X')
+B^{\mu\nu}(\partial_{\mu}X\partial_{\nu}X')\,,
\label{eq:totalD}
\eeqa
where the right-hand-side contains only two terms and a possible third term
vanishes,\\
$\,B^{\mu\nu}(X\partial_{\mu}\partial_{\nu}X')=0\,$,
because $B_{\mu\nu}$ is anti-symmetric.
The equation of motion for $B^{\mu}$ reads,
\begin{equation}
\partial_{\mu}B^{\mu\nu}
=-g' Y_{\ell_R}^{}\bar{\ell}_R\gamma^{\nu}\ell_R
 -g'Y_L^{}\bar{L}\gamma^{\nu}L
+\cdots \,,
\label{eq:DBmunu}
\end{equation}
where the coefficients $Y_{\ell_R}^{}$ and $Y_L^{}$ denote
the leptonic hypercharges of the right-handed singlet $\ell_R^{}$
and the left-handed doublet $L$\,, respectively.
Thus, from Eqs.\eqref{eq:totalD}-\eqref{eq:DBmunu},
we derive the following relations:
\beqs
\beqa
B^{\mu\nu}(\partial_{\mu}X\partial_{\nu}X')
&\!\!=\!\!& -\(\partial_{\mu}B^{\mu\nu}\)(X\partial_{\nu}X')
\nn
\\
&\!\!=\!\!& (X\partial_{\mu}X')\!
\left( g'Y_{\ell_R}\bar{\ell}_R\gamma^{\mu}\ell_R
+ g'Y_{L}\bar{L}\gamma^{\mu}L
\right)
+\cdots \,.
\\[1.5mm]
B^{\mu\nu}(\partial_{\mu}X\partial_{\nu}X')
&\!\!=\!\!& -B^{\mu\nu}(\partial_{\mu}X'\partial_{\nu}X)
\nn
\\
&\!\!=\!\!& -(X'\partial_{\mu}X)\!
\left( g'Y_{\ell_R}\bar{\ell}_R\gamma^{\mu}\ell_R
+ g'Y_{L}\bar{L}\gamma^{\mu}L
\right)
+\cdots \,.
\eeqa
\eeqs
Hence, we conclude that Eq.\eqref{eq:BXX} can be converted into
the operators \eqref{eq:LLXX-} and \eqref{eq:eeXX-}.

\vspace*{1mm}

In summary, the above proof demonstrates that
the set of operators in Eq.\eqref{eq:O6} are unique for our present EFT study.

\vspace*{7mm}
\noindent
{\bf\large Acknowledgements}
\\[1mm]
This research was supported in part by the National NSF of China
(under grants 11675086 and 11835005),
by the CAS Center for Excellence in Particle Physics (CCEPP),
by the National Key R\,\&\,D Program of China (No.\,2017YFA0402204),
by the Key Laboratory for Particle Physics, Astrophysics and Cosmology
(Ministry of Education),
and by the Office of Science and Technology, Shanghai Municipal Government
(No.\,16DZ2260200).

\baselineskip 17pt

\vspace{2mm}
%

\end{document}